\documentclass[12pt,useAMS,usenatbib,usegraphicx,iop]{emulateapj}
\usepackage{natbib,epsfig,graphicx,amssymb,color}
\usepackage[usenames,dvipsnames,svgnames,deluxetable]{xcolor}
\pdfoutput=1

\def\msun{{\rm M}_{\odot}}
\def\rsun{{\rm R}_{\odot}}
\def\lsun{{\rm L}_{\odot}}

\def\fluxcgs{10$^9$ erg s$^{-1}$ cm$^{-2}$}
\def\feh{{\left[{\rm Fe}/{\rm H}\right]}}
\def\fave{\langle F \rangle}
\def\bjdtdb{\ensuremath{\rm {BJD_{TDB}}}}

\def\mj{\ensuremath{\,{\rm M}_{\rm J}}}

\def\teff{{\rm T}_{\rm eff}}

\def\ch4{\rm CH_{4}}
\def\co2{\rm CO_{2}}
\def\h2o{\rm H_{2}O}
\begin{document}

\title{Discovery of water at high spectral resolution in the
  atmosphere of 51 P\lowercase{eg b}}

\author{J. L. Birkby\altaffilmark{1,2}}
\affil{Harvard-Smithsonian Center for Astrophysics,\\
60 Garden Street, Cambridge MA 02138, USA; and\\
Leiden Observatory, Leiden University, \\
Niels Bohrweg 2, 2333 CA Leiden, The Netherlands}

\author{R. J. de Kok}
\affil{Leiden Observatory, Leiden University, \\
Niels Bohrweg 2, 2333 CA Leiden, The Netherlands; and\\
SRON Netherlands Institute for Space Research,\\
Sorbonnelaan 2, 3584 CA Utrecht, The Netherlands}

\author{M. Brogi\altaffilmark{3}}
\affil{Center for Astrophysics and Space Astronomy,\\
University of Colorado at Boulder, Boulder, CO 80309, USA}

\author{H. Schwarz, I. A. G. Snellen}
\affil{Leiden Observatory, Leiden University,\\
Niels Bohrweg 2, 2333 CA Leiden, The Netherlands}

\altaffiltext{1}{NASA Sagan Fellow}
\altaffiltext{2}{jbirkby@cfa.harvard.edu}
\altaffiltext{3}{NASA Hubble Fellow}
\begin{abstract}
  We report the detection of water absorption features in the day side
  spectrum of the first-known hot Jupiter, 51 Peg b, confirming the
  star-planet system to be a double-lined spectroscopic binary. We
  used high-resolution ($R\approx100~000$), $3.2\mu$m spectra taken
  with CRIRES/VLT to trace the radial-velocity shift of the water
  features in the planet's day side atmosphere during $4$ hours of its
  4.23-day orbit after superior conjunction. We detect the signature
  of molecular absorption by water at a significance of $5.6\sigma$ at
  a systemic velocity of $V_{\rm sys}=-33\pm2$ km~s$^{-1}$, coincident
  with the 51 Peg host star, with a corresponding orbital velocity
  $K_{P} = 133^{+4.3}_{-3.5}$ km~s$^{-1}$. This translates directly to
  a planet mass of $M_{p}=0.476^{+0.032}_{-0.031}M_{\rm J}$, placing
  it at the transition boundary between Jovian and Neptunian
  worlds. We determine upper and lower limits on the orbital
  inclination of the system of $70^{\circ}<i<82.2^{\circ}$. We also
  provide an updated orbital solution for 51 Peg b, using an extensive
  set of 639 stellar radial velocities measured between 1994 and 2013,
  finding no significant evidence of an eccentric orbit. We find no
  evidence of significant absorption or emission from other major
  carbon-bearing molecules of the planet, including methane and carbon
  dioxide. The atmosphere is non-inverted in the temperature-pressure
  region probed by these observations. The deepest absorption lines
  reach an observed relative contrast of $0.9\times10^{-3}$ with
  respect to the host star continuum flux at an angular separation of
  3 milliarcseconds. This work is consistent with a previous tentative
  report of K-band molecular absorption for 51 Peg b by Brogi et
  al. (2013).
\end{abstract}

\section{Introduction}
The field of exoplanets has come of age, with twenty-one years passing
since the first confirmation of an exoplanet orbiting a main sequence
star, 51 Peg b \citep{May95}. The close, 4.23-day orbit of this planet
placed it in an entirely new and unexpected population of
highly-irradiated bodies close to their parent stars. It ignited the
field of planet migration theory \citep{Lin96,Ras96}, and paved the
way for another 3434 confirmed exoplanets in 2568 planetary systems to
date\footnote{As of 13 June 2016, see http://www.exoplanet.eu,
  \citealt{Sch11}.}. In just two decades, exoplanets have transitioned
from mere theoretical possibility to highly characterizable
systems. There are now radius measurements of Earth-like planets,
aided by asteroseismology, with error bars precise to $120$ km
\citep{Bal14}; there is evidence that clouds pervade the atmospheres
of exoplanets across the mass spectrum from super-Earths to hot
Jupiters (e.g. \citealt{Kre14a,Eva13,Sin16,Heng16}); there are a
growing number of robust detections of elemental and molecular species
in transiting planets using the Hubble Space Telescope, including
sodium, potassium and water (see e.g. \citealt{Cross15} for an
up-to-date review of chemicals observed in exoplanet atmospheres),
alongside the first detections of carbon monoxide, water, and methane
in the atmospheres of widely-separated directly imaged giants planets
(e.g. \citealt{Kono13,Sne14,Barm15,Maci15}), and in the atmospheres of
non-transiting hot Jupiters using ground-based high-resolution
spectroscopy \citep{Brog12,Rod12,Loc14,Brog14}. Even the global wind
dynamics and atmospheric circulation of hot Jupiters have been studied
in detail (e.g. \citealt{Knu09,Stev14,Lou15,Brog16,Zho16}). The next
few decades hold promise of remote, ground-based biomarker hunting in
Earth-like planets orbiting nearby bright stars
(e.g. \citealt{Sne13,Rod14,Sne15}), as well as the mapping of features
akin to Jupiter's Great Red Spot in the atmospheres of giant
exoplanets with the extremely large telescopes
\citep{Kos13,Sne14,Cross14,Kar15}, in a similar vein to that already
achieved for brown dwarfs \citep{Cross14map}. The discovery of 51 Peg
b was, in short, transformational.

However, its discovery was initally met with uncertainty and caution,
given its unusual orbital parameters. It was suggested that the radial
velocity measurements that revealed the planet were instead line
profile variations caused by non-radial stellar oscillations
\citep{Gray97,Gray97Hatz}. Although this claim was later retracted in
light of additional observations \citep{Gray98}, the rapid onslaught
of similar discoveries (e.g. \citealt{But97}), and the eventual
detection of transiting exoplanets \citep{Char00,Hen00}, largely laid
to rest any doubts about the planetary nature of the non-transiting
planet orbiting 51 Peg. As a final proof, in this paper, we
demonstrate the true binary nature of the 51 Peg star-planet system,
revealing it to be a double-lined spectroscopic (non-eclipsing)
binary, via the direct detection of water absorption lines in the
spectrum of the planet's atmosphere that undergo a change in
Doppler-shift.

The technique employed in this work uses ground-based, high-resolution
spectroscopy to directly observe the large radial velocity change
($\Delta$RV$_{\rm P}\sim$~km~s$^{-1}$) of the planet's spectrum while
the contamination from Earth's telluric features and the stellar lines
are essentially stationary ($\Delta$RV$_{\rm \star}\sim$~m~s$^{-1}$).
It works on the premise that at high resolution (e.g. $R\sim100~000$),
broad molecular bands are resolved into a dense forest of tens to
hundreds of individual lines in a pattern that is unique to each
molecule. Consequently, a significant correlation between a
high-resolution molecular template and the observed planetary
spectrum, at a systemic velocity that is coincident with the host
star, is evidence of the presence of a specific molecule in the
planet's atmosphere that is difficult to mimic with instrumental or
Earth-atmosphere systematics. The concept of using high-resolution
spectroscopy in this manner to study exoplanet atmospheres arose not
long after the discovery of 51 Peg b. \citet{Char98} initially
considered that reflected light from 51 Peg b may have been
responsible for the line profile variations proposed by
\citet{Gray97Hatz}, which lead to searches with high-resolution
optical spectroscopy to directly detect hot Jupiters via their
reflected light. \citet{Char99} and \citet{Col99} announced upper
limits and even a detection, respectively, of the reflected light from
$\tau$ Boo b, although ultimately they converged to an upper limit on
the star-planet flux ratio of $F_{p}/F_{s}<3.5\times10^{-5}$
\citep{Col04}. Many of the diagnostic properties of high-resolution
spectra of exoplanets were outlined by \citet{Bro01} and multiple
attempts followed to directly detect the thermally emitted light from
giant exoplanets at infrared wavelengths using high-resolution
spectrographs such as NIRSPEC on Keck II and Phoenix on Gemini South
(e.g. \citealt{Bro02,Dem05b,Barnes07b,Barnes07}), but again only
providing upper limits. It wasn't until \citet{Sne10} used the
CRyogenic high-resolution InfraRed Echelle Spectrograph (CRIRES) at
the Very Large Telescope (VLT) that the technique delivered its first
unambiguous detections of a molecule (carbon monoxide) in the
atmosphere of an exoplanet. While weather may have thwarted some
earlier attempts with other telescopes, the stability of CRIRES, in
part delivered by its use of adaptive optics and Nasmyth mounting, and
its higher spectral resolution, were undoubtedly instrumental to its
success. Since then, the technique has been used to study the
atmospheric composition of both transiting
\citep{Cross11,Bir13,deK13,Rod13,Schw15,Hoe15,Brog16} and
non-transiting exoplanets
\citep{Brog12,Rod12,Brog13,Brog14,Loc14,Sne14,
  Schw16,Pis16}. Additionally, the technique reveals the inclination,
$i$, of the orbit, thus the mass of the planet can be measured
directly in both cases, rather than just a lower $M_{\rm P}\sin(i)$
limit for the non-transiting planets.

The use of high-resolution infrared ground-based spectroscopy in this
paper further cements the important role of high-resolution optical
and infrared spectrographs in studying the atmospheres of
non-transiting planets. This is pertinent given that the nearest
potentially habitable, non-transiting, terrestrial planets orbiting
small stars (which have the most favourable contrast ratios) are
likely to be a factor of four times closer to Earth than their
transiting counterparts \citep{Dress14b}. Finally, this paper also
serves as an independent confirmation of the tentative detection of
molecular absorption features in 51 Peg b reported in
\citet{Brog13}. While we have used the same instrument as
\citet{Brog13} for our observations, we operated under a different
instrument setup to observe a redder wavelength region, and our method
for removing telluric contamination also differs, thus we consider our
results to be independent.

The paper is presented as follows: Section~\ref{sec:obs} describes our
observations, the instrumental set-up, and the reduction of the
resulting spectra of 51 Peg b, including the process of removing
telluric contamination. In Section~\ref{sec:results}, we detail the
cross-correlation process used to extract the planetary signal and
present our results. This includes an update to the orbital solution
and ephemeris for 51 Peg b.  Section~\ref{sec:discussion} presents a
discussion of our findings and compares them with preliminary reports
of molecular absorption in the atmosphere of 51 Peg b by
\citet{Brog13}. We conclude in Section~\ref{sec:conclusions}.

\section{Observations and Data Reduction}\label{sec:obs}
\subsection{Observations}
\begin{figure*}
\centering
\includegraphics[width=\textwidth]{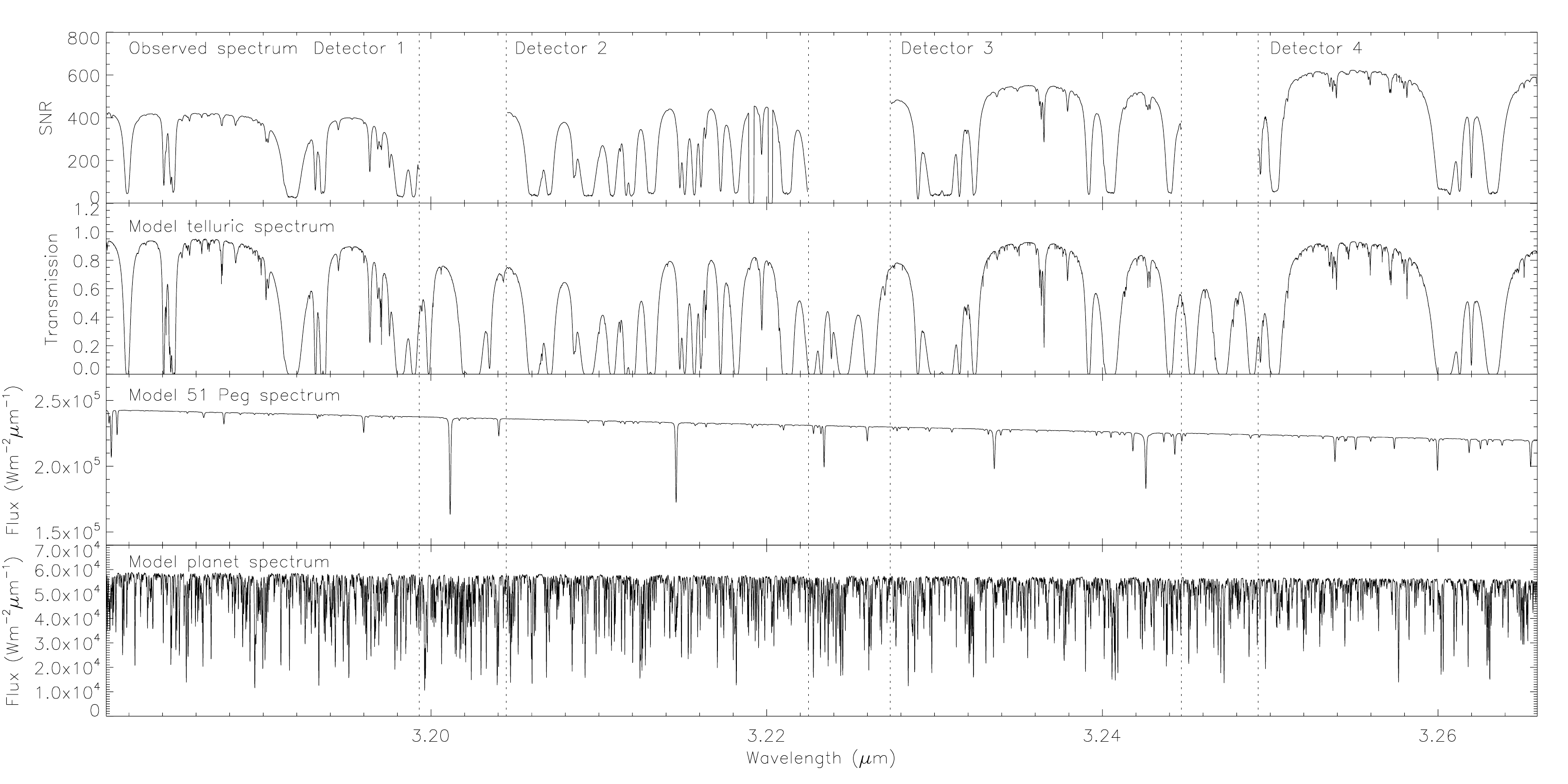}
\caption{\textbf{Top:} The photon-limited average signal-to-noise of
  the 51 Peg spectra observed with CRIRES/VLT. The vertical dotted
  lines mark of boundaries of the gaps between the
  detectors. \textbf{Second panel:} A model telluric transmission
  spectrum from ATRAN assuming a precipitable water vapour PWV=2 mm at
  Cerro Paranal. The observed spectra are completely dominated by the
  tellurics. \textbf{Third panel:} For visual purposes only, an
  approximate stellar model for 51 Peg, assumed here to be the Solar
  spectrum, shifted to match the velocity of 51 Peg during our
  observations. The spectrum was obtained at $R=100~000$. Most of the
  strong stellar lines fall between the detector gaps. \textbf{Bottom
    panel:} An example of one of our water molecular template spectra
  for 51 Peg b (see Section~\ref{sec:models}), shifted to the velocity
  of 51 Peg during our observations. Note the many tens of strong
  absorption lines.}
\label{fig:spectrae}
\end{figure*}

We observed the bright star 51 Peg (G2.5V, $V = 5.46$ mag, $K = 3.91$
mag) for $3.7$ hours during the night beginning 2010 October 21, using
CRIRES\footnote{CRIRES was dismounted from UT1/VLT in the summer of
  2014 to be upgraded to CRIRES$+$, which will have improved detectors
  and a wider wavelength coverage \citep{Fol14}. Its return is eagerly
  anticipated.} ; \citep{Kae04} mounted at Nasmyth A at the Very Large
Telescope (VLT, 8.2-m UT1/Antu), Cerro Paranal, Chile. The
observations were collected as part of the ESO large program
186.C-0289. The instrument setup consisted of a 0.2 arcsec slit
centred on 3236 nm (order 17), in combination with the
Multi-Application Curvature Adaptive Optic system (MACAO;
\citealt{Ars03}). The CRIRES infrared detector is comprised of four
Aladdin III InSb-arrays, each with $1024\times512$ pixels, and
separated by a gap of $280$ pixels. The resulting wavelength coverage
of the observations was $3.1806<\lambda(\mu$m$)<3.2659$ with a
resolution of $R\approx100~000$ per resolution element (see
Figure~\ref{fig:spectrae}).

We observed 51 Peg continuously while its hot Jupiter companion passed
through orbital phases $0.55\lesssim\phi\lesssim0.58$, corresponding
to an expected change in the planet's radial velocity of
$\Delta$RV$_{\rm P}=-23$ km~s$^{-1}$ (15 pixels on the CRIRES
detectors). In total, we obtained $42$ spectra, with the first 20
spectra each consisting of two sets of $5\times20$ second exposures,
and the remainder each consisting of two sets of $5\times30$ second
exposures. The increase in the exposure time was aimed at maintaining
a constant signal-to-noise ratio (SNR) in the continuum of the
observed stellar spectra after a sudden and significant deterioration
of the seeing (increasing from $0.75$ arcsec to $1.4$ arcsec between
one set of frames, see Section~\ref{sec:sysrem}). To enable accurate
sky-background subtraction, the telescope was nodded along the slit by
10 arcsec between each set of exposures in a classic ABBA sequence,
with each of the final 42 extracted spectra consisting of an AB or BA
pair. A standard set of CRIRES calibration frames was taken the
following morning. Later in this paper, we will compare our results to
those of \citet{Brog13}, who observed 51 Peg b at $2.3~\mu$m at the
same spectral resolution with CRIRES/VLT on dates either side of these
observations, including 2010 October 16, 17, and 25.

\subsection{Data Reduction}\label{sec:reduction}
Throughout the data reduction, the four CRIRES detectors were treated
independently and separately. We used the CRIRES \textsc{esorex}
pipeline (v2.2.1) to first process the observed 2-D images, including
non-linearity and bad pixel corrections, flat-fielding, background
subtraction and combination of the nodded exposures, and finally the
optimal extraction of the 1-D spectra. The 42 extracted spectra were
stored as four $1024\times42$ matrices. The matrix for detector 1 is
shown in the upper panel of Figure~\ref{fig:sysrem}. The x-axis
corresponds to pixel number (i.e.  wavelength channel) and the y-axis
denotes the frame number (i.e.  orbital phase or time). Remaining
singular bad pixels, bad regions, and bad columns in these matrices
were identified iteratively by eye and replaced by spline
interpolation values from their horizontal neighbouring pixels. There
was a total of $0.4-1.0$ per cent bad pixels in each matrix, with
detector 4 requiring the most corrections.

A gradual drift occurs in the position of the spectrum in the
dispersion direction on the detectors over the course of the
observations. To correct this, we apply a global shift to each
spectrum on each detector using a spline interpolation to align it to
the telluric features of the spectrum with the highest SNR. The shifts
were determined by cross-correlating the spectrum in question with the
highest SNR spectrum using \textsc{iraf.fxcor}. Detector 1 required
the largest corrections, with last spectrum deviating from the first
by $0.7$ pixels (the equivalent of $1$ km~s$^{-1}$ or
$0.01$\AA). \citet{Brog13} note that their $2.3~\mu$m observations of
51 Peg b also experience a similar drift, which correlates with the
temperatures of the instrument pre-optics system, grating and its
stabilizer. For small fluctuations ($<0.05$ K), the drift did not
exceed 0.5 pixels, but a $1.5$ K change in these temperatures resulted
in much larger drifts ($1.5$ pixels) for their observations on 2010
October 25, which resulted in a non-detection of the 51 Peg b
signal. The drift in our $3.5~\mu$m observations does not correlate
with these instrumental temperatures, which remained stable throughout
the night. Instead, the drift correlates with the ambient temperature
of the telescope dome and the primary mirror temperature, which both
cooled by $2$ K over the course of the observations. However, the
drift of our $3.5~\mu$m spectra is comparatively small, thus we do not
expect the alignment correction to significantly affect our subsequent
analysis.

Finally, a common wavelength solution per detector was calculated
using a synthetic telluric transmission spectrum (see the second panel
of Figure~\ref{fig:spectrae}) from
ATRAN\footnote{http://atran.sofia.usra.edu/cgi-bin/atran/atran.cgi}
\citep{Lor92} to identify the wavelengths of the telluric features in
the highest SNR spectrum. Line positions were identified using
\textsc{iraf.identify}, and fitted with a third order Chebyshev
polynomial to obtain the wavelength solution. This replaced the
default solution from the CRIRES pipeline from which it differed by up
to $1.9$\AA. The wavelength solutions were not linearized and thus
retained the pixel-spacing information. Detector 2 contains
significant telluric contamination (see Figures~\ref{fig:spectrae}
and~\ref{fig:sysrem}) such that no useful planetary signal can be
extracted. Following \citet{Bir13}, we therefore discarded detector 2
and exclude it from all further analyses in this paper.

\subsection{Removal of telluric contamination}\label{sec:sysrem}
Telluric absorption from the Earth's atmosphere is the dominant
spectral feature in our observed spectra (see
Figure~\ref{fig:spectrae}), while the Doppler-shifted features of 51
Peg b are expected at the $10^{-3}-10^{-4}$ level with respect to the
stellar continuum. Thus, we needed to remove the telluric
contamination. In a previous analysis of high-resolution spectra of 51
Peg b, \citet{Brog13} included an additional step to remove stellar
lines before removing the telluric features. However, they studied the
$2.3\mu$m region, which contains multiple strong CO lines from the
Sun-like host star. In the $3.2\mu$m region under consideration here,
a comparison with a proxy solar model spectrum for 51 Peg from Robert
Kurucz's stellar model
database\footnote{http://kurucz.harvard.edu/stars/sun/} at
$R=100~000$ indicates that there are few strong absorption lines from
the host star, and that they mostly fall on gaps between detectors, or
on the discarded detector 2 (see the third panel of
Figure~\ref{fig:spectrae}).  Consequently, we do not perform any
pre-removal of the stellar lines in the $3.2\mu$m data set.

\begin{figure*}
\centering
\includegraphics[trim=1cm 0cm 1cm 1.25cm, width=\textwidth]{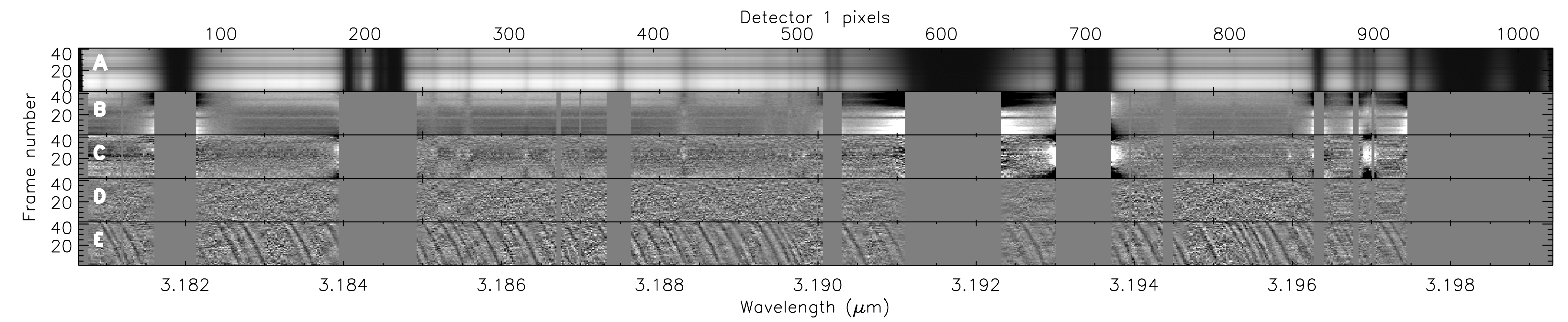}
\includegraphics[trim=1cm 0cm 1cm 0cm, width=\textwidth]{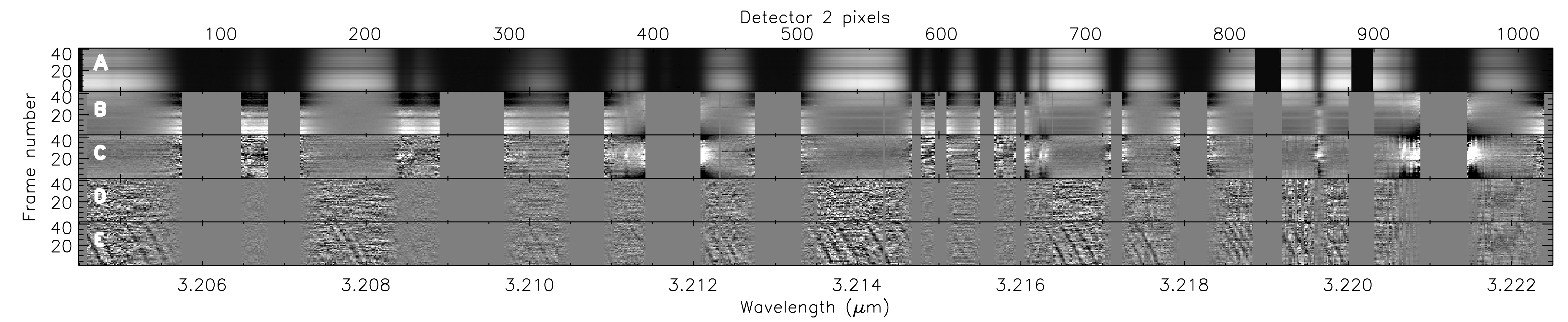}
\includegraphics[trim=1cm 0cm 1cm 0cm, width=\textwidth]{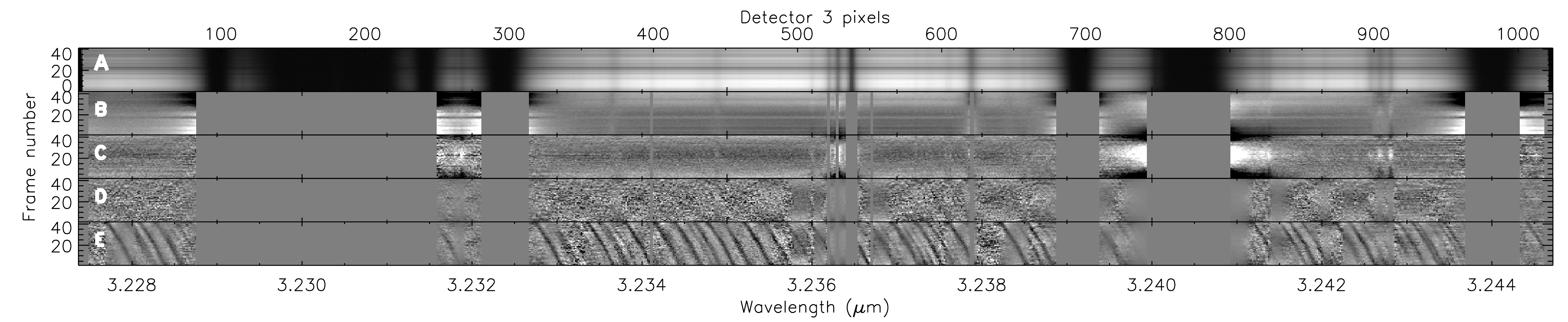}
\includegraphics[trim=1cm 0cm 1cm 0cm, width=\textwidth]{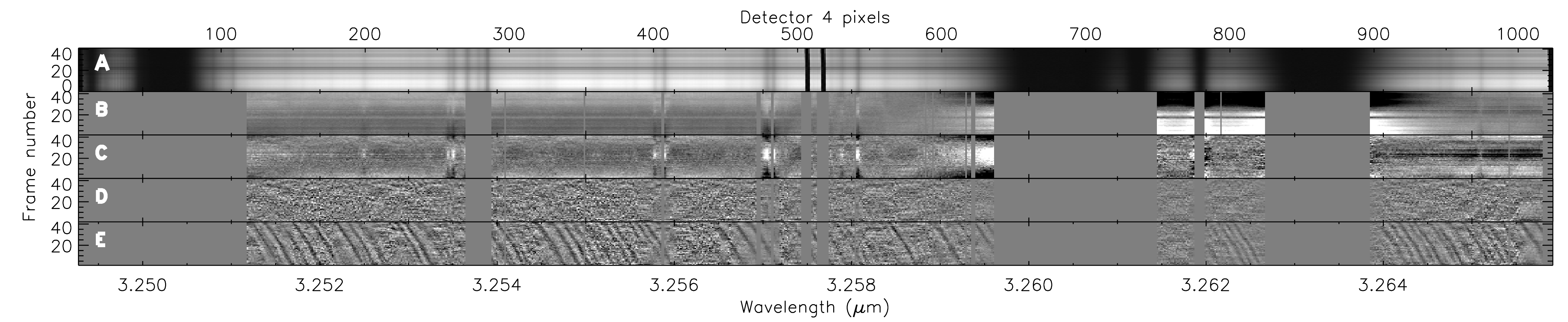}
\caption{\scriptsize{Spectra at different stages of the telluric
    removal process for each CRIRES detector. The x-axes correspond to
    wavelength i.e. pixel number, and the y-axes are ordered in time
    i.e. frame number. Detector 2 is not used in our analysis but it
    shown here for completeness. The sub-panels are as
    follows. \textbf{Panel A:} The spectra extracted from the CRIRES
    pipeline, with bad pixels corrected, and aligned to match the
    telluric features of the highest SNR spectrum. The dark horizontal
    bands contain spectra taken under poor seeing. The broad dark
    vertical bands are saturated telluric lines. \textbf{Panel B:} As
    in A, but normalised and with the mean of each column subtracted
    from itself. The solid grey regions mark regions of saturated
    telluric features which are excluded from our cross-correlation
    analysis (see Section~\ref{sec:xcor}).  \textbf{Panel C:} The
    residuals remaining after one iteration of \textsc{Sysrem} on the
    spectra. Note that the non-saturated telluric and stellar features
    e.g. at pixel $420$ on detector 1 still remain.  \textbf{Panel D:}
    The residuals after applying the adopted number of iterations of
    \textsc{Sysrem} for the detector, the high-pass filter, and
    dividing each column by its variance . The telluric features have
    been sufficiently removed, leaving behind the planet spectrum
    buried in the noise.  \textbf{Panel E:} The same as D, but with a
    best-matching model planet spectrum from Section~\ref{sec:xcor}
    injected at the expected Doppler shift of 51 Peg b at a factor of
    100 times greater than its nominal value before running
    \textsc{Sysrem}. This is to highlight the many individual strong
    water lines in the planet spectrum whose signal will be combined
    with the cross-correlation procedure detailed in
    Section~\ref{sec:xcor}. The authors are happy to supply the
    processed spectral matrices upon request.}}
 \label{fig:sysrem}
\end{figure*}
 
\begin{figure}
\centering
\includegraphics[trim=1.2cm 0cm 0cm 0cm, width=0.5\textwidth]{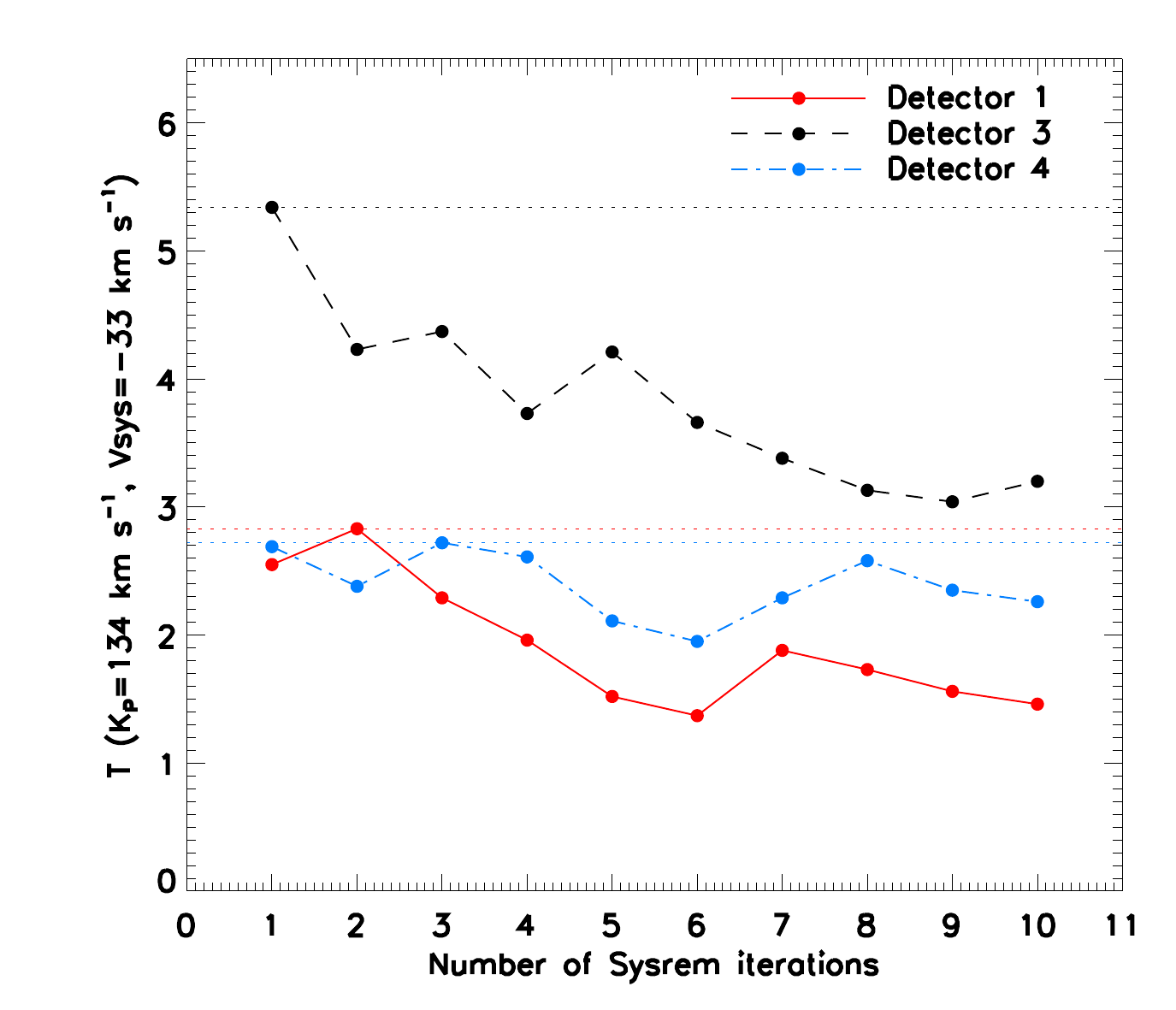}
\caption{The detection strength (T, as described in
  Section~\ref{sec:significance}) of an injected fake planet at the
  proposed planet velocity parameters from \citet{Brog13} for each
  CRIRES detector after each iteration of \textsc{Sysrem}. The
  horizontal dotted lines mark the maximum detection strength per
  detector for the injected model.}
\label{fig:sysrem_iters}
\end{figure}

The removal of the telluric features in our spectra was achieved using
our implementation of \textsc{Sysrem}, which is an algorithm based on
Principle Component Analysis (PCA) but also allows for unequal error
bars per data point \citep{Tam05,Maz07}. It is commonly used in
ground-based wide-field transit surveys to correct systematic effects
common to all light curves (e.g.~SuperWASP; \citealt{Col06}). Each
wavelength channel (i.e. pixel column) in the matrix shown in panel B
of Figure~\ref{fig:sysrem} was treated as a `light curve', where the
errors per data point are the quadrature sum of the Poisson noise and
the error from the optimal extraction of the spectrum at a given
pixel. Each column had its mean subtracted before being passed through
\textsc{Sysrem}, and regions of saturated telluric features, which
contain essentially no flux, were also masked. These masked regions
are marked by the vertical solid grey regions in
Figure~\ref{fig:sysrem}. \textsc{Sysrem} then searched for common
modes between the 1024 light curves per matrix, such as variation with
airmass, and subtracted them resulting in the removal of the
quasi-static telluric and stellar lines, leaving only the
Doppler-shifting planet spectrum in each spectrum plus noise. However,
in practice, once the dominant telluric and stellar spectral features
are removed, \textsc{Sysrem} will begin to remove the planet features
too. This is because the sub-pixel shift of the planet spectrum
between frames creates a small but detectable common mode between
adjacent columns. Thus, we must determine when to halt the
\textsc{Sysrem} algorithm before it removes the planet signal. To do
this, we injected the best-matching model planet spectrum proposed by
\citet{Brog13} based on observations at $2.3~\mu$m at their measured
planet velocity and ephemeris ($V_{\rm sys}=-33$ km~s$^{-1}$,
$K_{P}=134$ km~s$^{-1}$, with a phase shift of $\Delta\phi=0.0095$),
and iterated \textsc{Sysrem} ten times. The model was injected at a
nominal strength of $1$. At each iteration, we used the
cross-correlation method described in Sections~\ref{sec:xcor}
and~\ref{sec:significance} to determine the significance of the
detection at the injected velocity. The results of this analysis are
shown in Figure~\ref{fig:sysrem_iters} for each detector. We find that
\textsc{Sysrem} begins to remove the planet signal after only one
iteration on detector 3, but that this does not occur until after two
and three iterations on detectors 1 and 4, respectively. The
difference in iterations required per chip may be partly due to a
known `odd-even'
effect,\footnote{http://www.eso.org/sci/facilities/paranal/instruments/crires/doc/VLT-MAN-ESO-14500-3486\_v93.pdf}
which only affects detectors 1 and 4. It is caused by variations in
the gain between neighbouring columns along the spectral
direction. This odd-even effect has been seen previously in similar
analyses of high-resolution spectroscopy of exoplanets from CRIRES/VLT
\citep{Bir13,Brog12,Brog13,Brog14,Brog16}.

\begin{figure}
\centering
\includegraphics[trim=1.2cm 0cm 0cm 2cm, width=0.5\textwidth]{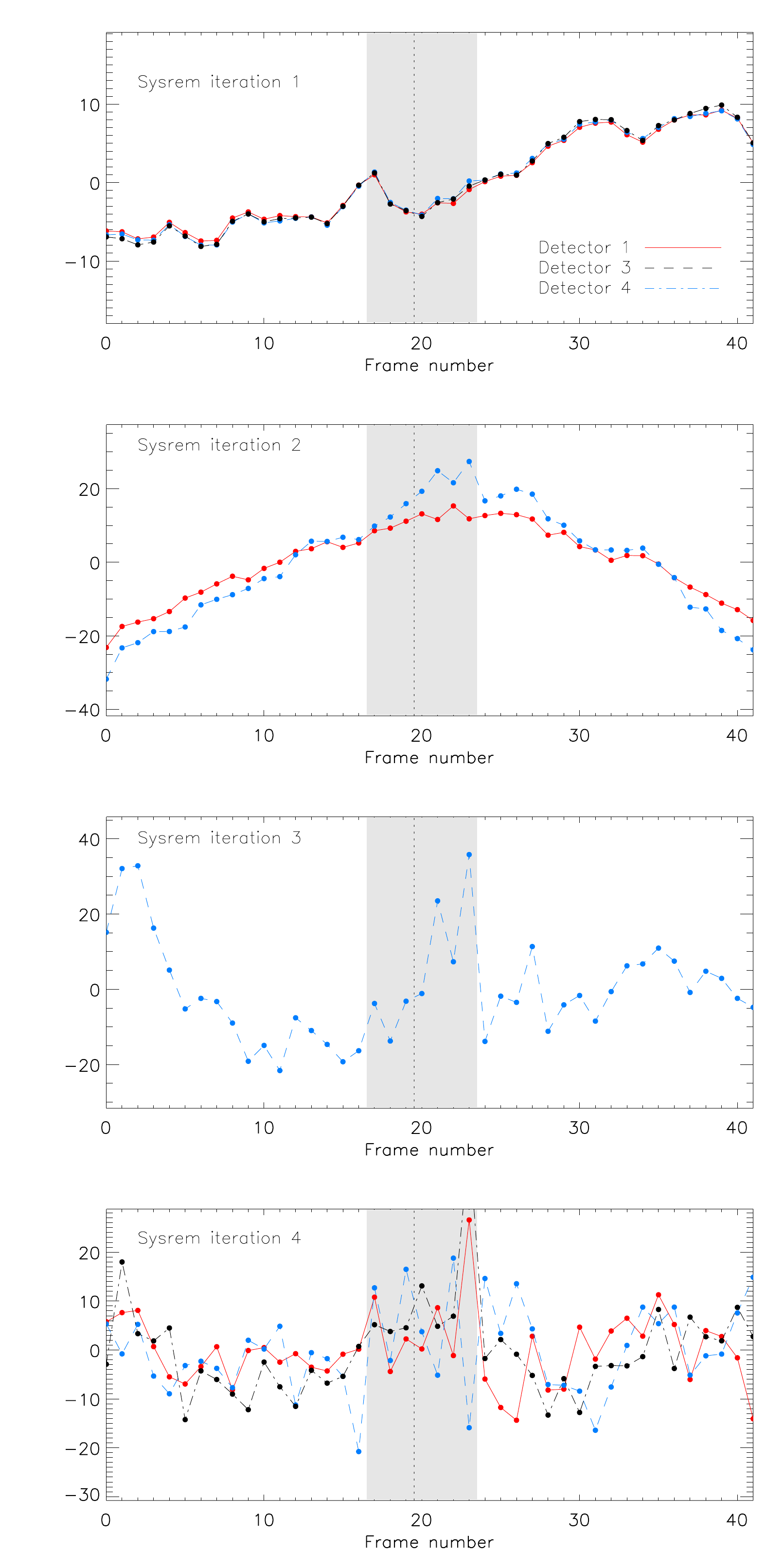}
\caption{The trend identified and removed from each column by
  \textsc{Sysrem} for each detector. Although it is not used in our
  final data analysis, the fourth \textsc{Sysrem} iteration for all
  detectors is shown in the bottom panel. Its purpose here is to
  highlight the overall relative flatness of the removed trend, except
  in the grey vertical regions which bound spectra that are later
  excluded from the analysis of the planetary signal (see
  Section~\ref{sec:degraded}). These spectra occurred during a period
  of poor and unsettled seeing (see Figure~\ref{fig:physical}). The
  vertical dashed line marks when the exposure time was increased from
  $20$ seconds to $30$ seconds.}
\label{fig:correlations} \end{figure}

\begin{figure}
\centering
\includegraphics[trim=0cm 0cm 0cm 1cm, width=0.5\textwidth]{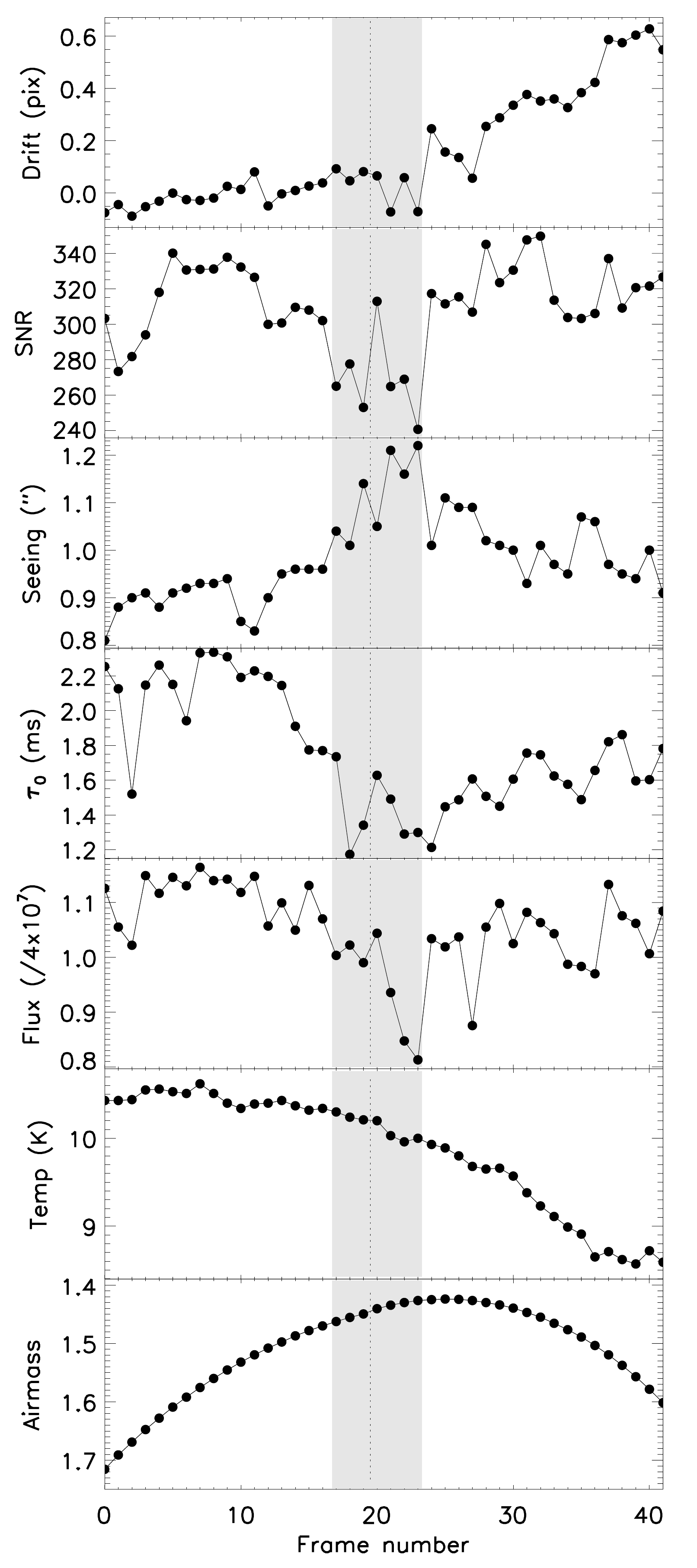}
\caption{Possible physical causes of the trends removed by
  \textsc{Sysrem}. The grey regions and dashed vertical lines are the
  same as for Figure~\ref{fig:correlations}. Note the sudden drop in
  SNR as the seeing begins to deteriorate. It recovers as the exposure
  time is increased but continues to degrade as the seeing worsens,
  only recovering when the seeing stabilises. This trend is coincident
  with a rapid change in the wavefront coherence time, $\tau_{0}$. The
  seeing and $\tau{0}$ were acquired from the VLT Astronomical Site
  Monitor (VLT-ASM). The flux is the raw value recorded by the
  adaptive optics (AO) sensor, and the temperature is the telescope
  ambient temperature.}
\label{fig:physical}
\end{figure}

Guided by the results of executing \textsc{Sysrem} on the injected
signal as described above, we adopted \textsc{Sysrem} iterations of
two, one, and three for detectors 1, 3, and 4, respectively, for the
analysis of the observed data (see Section~\ref{sec:xcor}). The
standard deviation of the final residuals in panels D of
Figure~\ref{fig:sysrem} are 0.0050, 0.038 0.0082, 0.0040, for
detectors 1, 2, 3, and 4, respectively. The trends removed from each
detector during each \textsc{Sysrem} iteration are shown in
Figure~\ref{fig:correlations}. Possible physical causes of these
trends are shown in Figure~\ref{fig:physical}, and these are discussed
in further detail in Section~\ref{sec:degraded}.

The final two steps in the telluric removal process were: i) the
application of a high-pass filter with a 64-pixel width smoothing
function, which removes a heavily smoothed version of each residual
spectrum from itself to filter out low-order variation across the
matrix, and then ii) each column is divided by its variance to account
for variation in SNR as a function of wavelength. The final product of
this process for detector 1 is shown in panel D of
Figure~\ref{fig:sysrem}. For illustrative purposes, the bottom panel
of Figure~\ref{fig:sysrem} shows how the final matrix would appear if
a model planet was injected at the planet velocity but at $100\times$
nominal value before running \textsc{Sysrem}. Many individual lines
from the planet spectrum are clearly visible as they blue-shift across
the matrix.  With the tellurics and stellar continuum removed from
each detector, we could proceed to extract the observed planetary
signal contained within the noise of the residual spectra.

\section{Cross-correlation Analysis and Results}\label{sec:results}
The spectral features of the molecules in the planetary atmosphere are
buried in the noise of the residuals after removing the telluric
contamination. To identify them, the signal from all the individual
spectral lines are combined by cross-correlating the residuals with
high-resolution molecular spectral templates; a form of chemical
`fingerprinting'. We searched the atmosphere of 51 Peg b for molecular
features arising from the expected major carbon- and oxygen-bearing
gases at the observed wavelengths, namely water (H$_{2}$O), carbon
dioxide (CO$_{2}$), and methane (CH$_{4}$), using a grid of model
atmospheres in a similar manner to \citet{Bir13} and
\citet{deK13}. Spectral features from carbon monoxide (CO) are not
expected in the observed $3.2~\mu$m region.

\subsection{Models}\label{sec:models}
The molecular templates are parametrized by a grid of atmospheric
temperature-pressure ($T-P$) profiles and trace gas abundances (i.e.
volume mixing ratios; VMR). To generate the model spectra, we employed
the same radiative transfer code of \citet{deK13}, performing
line-by-line calculations, including H$_{2}$-H$_{2}$ collision-induced
absorption along with absorption from the trace gases, which is
assumed to follow a Voigt line shape. We used line data from HITEMP
2010 \citep{Roth10} to create the H$_{2}$O and CO$_{2}$ models, and
HITRAN 2008 \citep{Roth09} for CH$_{4}$. The model atmospheres are
clear (i.e. cloud-free) with uniformly mixed gases. The $T-P$
structure follows a relatively simple profile. Deep in the atmosphere,
at pressures $p_{1}$ and higher, we assume a uniform $T-P$ profile at
a fixed temperature $t_{1}$. Between pressures $p_{1}$ and $p_{2}$, we
assume a constant lapse rate (i.e. a constant rate of change of
temperature with log pressure). At altitudes higher (and pressures
lower) than $p_{2}$ we again assume a uniform $T-P$ profile at fixed
temperature $t_{2}$. The pressure $p_{1}$ took values of (1, 0.1,
0.01) bar, and $p_{2}$ was varied using values of
($1\times10^{-3}, 1\times10^{-4}, 1\times10^{-5}$) bar.  The basal
temperatures were guided by the effective temperature of the planet
assuming external heating (see Equation 1 of \citealt{Lop07}), a low
Bond albedo ($A_{B}<0.5$), and considering the full range of heat
circulation from instantaneous reradiation to full advection. Thus,
$t_{1}$ took values of ($1000$, $1250$, $1500$) K, while $t_{2}$ was
varied using values of ($500$, $1500$) K. Note that certain
combinations result in inverted $T-P$ profiles, where the temperature
increases with increasing altitude (decreasing pressure). These
spectra have features in emission, rather than absorption. The gas
abundance volume mixing ratios took values appropriate for hot
Jupiters over our considered temperature range (e.g.
\citealt{Mad12}), including $10^{-4.5}$, $10^{-4}$, or $10^{-3.5}$ for
water, and $10^{-7}$, $10^{-5}$, or $10^{-3}$ for CO$_{2}$ and
CH$_{4}$.

Before cross-correlating the residuals with the molecular template
grid, we convolved the models to the spectral resolution of CRIRES,
and subtracted their baseline level. Note that the telluric removal
process in Section~\ref{sec:sysrem} has also removed the continuum
information in the observed planet spectrum, such that our analysis is
only sensitive to the relative, not absolute, depth of the spectral
features with respect to the stellar continuum.

\subsection{Cross-correlation analysis and results}\label{sec:xcor}
\begin{figure} \centering
  \includegraphics[width=0.5\textwidth]{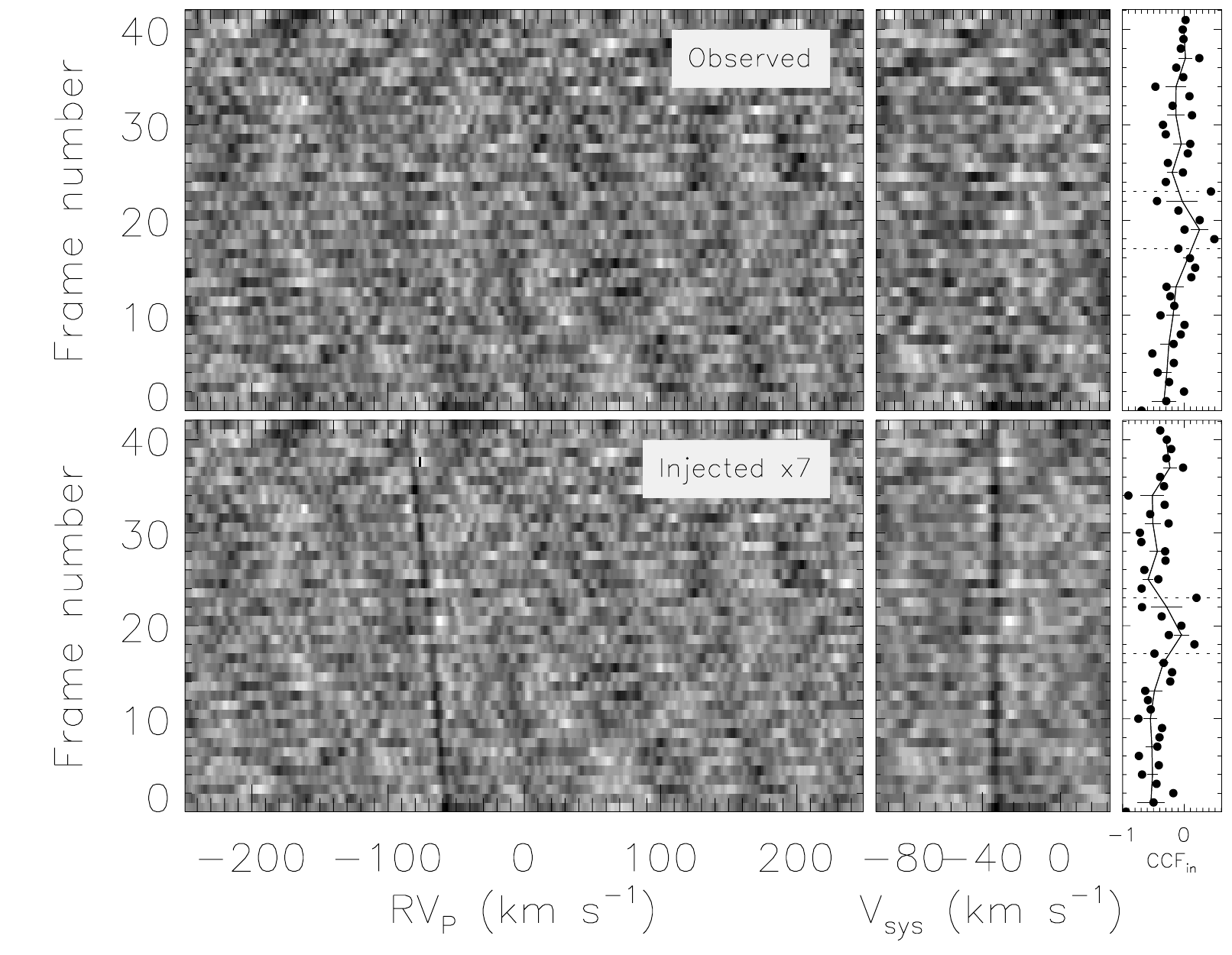}
  \caption{Cross-correlation functions for each spectrum using the
    best-matching model. \textbf{Top-left:} The summed CCFs for each
    residual spectrum after cross-correlating the best matching
    H$_{2}$O template with the observed data. \textbf{Bottom-left:}
    Same as top-left, but with a model spectrum injected at $7\times$
    its nominal value. Note the dark diagonal blue-shifting trail of
    the injected planet signal. \textbf{Top-middle:} As in top left
    panel but aligned into the rest frame of the planet. The trail is
    located at the known systemic velocity of the 51 Peg star system
    ($V_{\rm sys}=-33.25$ km~s$^{-1}$). \textbf{Bottom-middle:} Same
    as top-middle, but with the model injected. The \textbf{top-right}
    and \textbf{bottom-right} panels show the strength of the CCFs for
    a 3-pixel column bin centered on $V_{\rm sys}=-33$ km~s$^{-1}$
    (i.e. containing the planet signal, CCF$_{\rm in}$). The solid
    line shows the mean of 3-frame binning.  Note how the strength of
    the CCFs approach zero between frames $18-24$ (corresponding to
    phase$\approx0.565$). This occurs in both the observed spectra and
    in those where the model was injected prior to removing the
    telluric and stellar lines with \textsc{Sysrem}, and corresponds
    to the period of poor atmospheric conditions (see
    Figure~\ref{fig:physical}). These CCFs are removed from our final
    analysis in Section~\ref{sec:significance}, and are further
    detailed in Section~\ref{sec:degraded}.}
  \label{fig:trail}
\end{figure}

The cross-correlation analysis was performed for planet radial
velocities in the range $-249\leq\rm RV_{P} (\rm km~s^{-1})\leq249$ in
intervals of $1.5$ km~s$^{-1}$, interpolating the convolved grid of
molecular templates onto the Doppler-shifted wavelengths. The interval
size is set by the velocity resolution of the CRIRES pixels. The
cross-correlation functions (CCFs) were determined separately for each
residual spectrum on each detector, and then summed equally with their
corresponding CCF on the other detectors, resulting in a single summed
CCF matrix of dimension $333\times42$. The matrix created by the
best-matching template is shown in upper left panel of
Figure~\ref{fig:trail}, where the template is a water-only model with
the following parameters: $t_{1}=1500$ K, $t_{2}=500$K, $p_{1}=0.1$
bar, $p_{2}=1\times10^{-5}$ bar, and a water volume mixing ratio of
VMR$_{\rm H2O}=10^{-4}$. These parameters differ to the best-matching
model reported by \citet{Brog13} for molecular absoprtion at
$2.3~\mu$m; however, see Sections~\ref{sec:significance}
\&~\ref{sec:discussion} for further discussion on other models in the
grid that produce signals within $1\sigma$ of this result. We refer
the reader to Section~\ref{sec:degraded} for a more detailed
discussion of Figure~\ref{fig:trail}.

We note that in our first attempt to perform the cross-correlation
analysis of this data, we used the ephemeris and orbital solution for
51 Peg b in \citet{But06}, and refrained from using the phase shift
invoked by \citet{Brog13} to match the planet signal to the systemic
velocity of the host star, even though their $\Delta_\phi=0.0095$
phase shift was within the $1\sigma$ uncertainty range
$\Delta\phi=\pm0.012$ of the \citet{But06} ephemeris. However, we also
found that this resulted in the strongest cross-correlation signal
being offset from the known systemic velocity of the host star by
$-9\pm2$ km~s$^{-1}$, corresponding to a phase shift of
$\Delta\phi=0.011$. This is still within the uncertainty of the
original \citet{But06} orbital solution. However, since the discovery
of 51 Peg b, the RV of its host star has been monitored sporadically
throughout the decades, hence we endeavoured to measure the most
up-to-date ephemeris for the planets orbital solution in the hope that
this would negate the need for the phase shift in our analysis.

\subsubsection{A refined orbital solution for 51 Peg b}\label{sec:orbit}
To ensure we had its most up-to-date orbital solution, we compiled an
extensive repository of literature and archival radial velocity
measurements of the star 51 Peg from multiple observatories. These
data are given in Table~\ref{tab:rvdata} and span observing dates from
15 September 1994 to 9 July 2014, resulting in 639 RV measurements
over 20 years. The table includes the discovery measurements from the
ELODIE spectrograph at Observatoire Haute Provence \citep{May95} and
subsequent additional monitoring. We took these RV measurements from
the \citet{Nae04} compilation. We also included the legacy dataset
from Lick Observatory observed with the Hamilton spectrograph, taking
measurements from the self-consistent reprocessing of all the Lick
spectra presented by \citet{Fis14}. Finally, we included more recent
additional monitoring from HIRES at the Keck Observatory
\citep{How16}, and extracted RVs from observations with HARPS at the
ESO-3.6m telescope in 2013 (ESO program ID 091.C-0271, PI:
Santos). The reduced HARPS spectra were obtained from the ESO Science
Archive\footnote{http://archive.eso.org/wdb/wdb/adp/phase3\_spectral/query},
and the RVs, their errors, and timestamps were obtained from the
headers of the CCF data product files, using keywords \textsc{drs ccf
  rvc}, \textsc{drs dvrms}, and \textsc{drs bjd}, respectively. We
were careful to note the format of the timestamps reported for all our
datasets, which vary between JD, HJD, and BJD, due to different
conventions being adopted over time. The timestamps in
Table~\ref{tab:rvdata} were all converted to BJD$_{\rm TDB}$
(Barycentric Dynamical Time).

\begin{deluxetable}{llll}
\tablecaption{Stellar radial velocity measurements of 51 Peg from
  multiple observatories}
\tablehead{\colhead{BJD$_{\rm TDB}$}&\colhead{RV (m~s$^{-1}$)}&\colhead{$\sigma_{\rm RV}$ (m~s$^{-1}$)}&\colhead{Dataset}}
  \startdata
2449610.532755&   -33258.0&        9.0&   ELODIE\\
2449612.471656&   -33225.0&        9.0&   ELODIE\\
2449655.311263&   -33272.0&        7.0&   ELODIE\\
...&   ...&        ...&   ...
  \enddata
  \tablecomments{The Lick8 RVs reported in this table include a
    $+13.1$ m~s$^{-1}$ velocity offset correction to the RVs extracted
    from
    Vizier\footnote{\scriptsize\mbox{http://vizier.cfa.harvard.edu/viz-bin/VizieR-3?-source=J/ApJS/210/5/table2}},
      to account for the instrumental systematic reported in
      \citet{Fis14}. The following additional offsets, determined from
      a circular orbit fit to each dataset using \textsc{Exofast}, can
      be applied to place all of the RV measurements onto the same
      zero-point: $\gamma_{\rm~Lick13}=-21.70$ m~s$^{-1}$,
      $\gamma_{\rm~Lick8}=-4.52$ m~s$^{-1}$,
      $\gamma_{\rm~Lick6}=-14.64$ m~s$^{-1}$,
      $\gamma_{\rm~ELODIE}=+33251.59$ m~s$^{-1}$,
      $\gamma_{\rm~HIRES}=+2.24$ m~s$^{-1}$,
      $\gamma_{\rm~HARPS}=+33152.54$
      m~s$^{-1}$. Table~\ref{tab:rvdata} is published in its entirety
      in the machine-readable format.  A portion is shown here for
      guidance regarding its form and content.}
\label{tab:rvdata}
\end{deluxetable}

We used
\textsc{Exofast}\footnote{http://astroutils.astronomy.ohio$-$state.edu/exofast/exofast.shtml}
\citep{Eas13} to model the orbital components constrained by each
dataset. To find the best fitting model to the radial velocity data,
\textsc{Exofast} employs a non-linear solver (\textsc{amoeba}), which
uses a downhill simplex to explore the parameter space that minimises
the $\chi^{2}$ of the orbital solution. In order to negate any
systematic underestimate of the uncertainties on the RV data,
\textsc{Exofast} rescales the RV uncertainties by a constant
multiplicative factor, such that the reduced $\chi^{2}$ of the best
fit is unity ($\chi^{2}_{\nu}=1$). Consequently, a poor fit would be
reflected by larger uncertainties on the derived
parameters. \textsc{Exofast} does not include an additive jitter term
to the RV solution, as previous work with \textsc{Exofast} found no
statistically significant difference between the two approaches to
uncertainty scaling (e.g. \citealt{Lee11}). Once the best fit is
found, \textsc{Exofast} then executes a differential evolution Markov
chain Monte Carlo method to obtain the uncertainties on the derived
orbital elements. These algorithms are explained in detail in
\citet{Eas13}. The code requires priors for the stellar effective
temperature ($\teff=5787\pm233$ K), metallicity
($\feh=0.200\pm0.030$), and surface gravity
($\log(g_*)=4.449\pm0.060$), which we supplied based on
\citet{Val05}. We also gave the logarithm of the period
($\log_{10}(P)=\log_{10}(4.230785\pm0.000036)$) as a prior based on
\citet{But06}, and restricted the period range to $4-5$ days.

Combining relative RV measurements from different observatories is
hindered by velocity offsets in part due to instrumental
systematics. We consider our repository to consist of six separate
datasets: HARPS, HIRES, ELODIE, plus three Lick datasets (Lick13,
Lick8, and Lick6). The numbers in the Lick dataset names denote the
dewar associated with each upgrade to the Hamilton spectrograph, which
introduced different velocity offsets (see \citealt{Fis14} for further
information). Our first step was to model the datasets independently
to determine if they supported an eccentric orbit. We ran two models,
one with and one without a free variable for a long-term linear trend
in the RVs. We exclude the HARPS dataset in this assessment, as it has
poor phase coverage, being clustered within $\phi=\pm0.1$ of superior
and inferior conjunction and thus lacking in strong constraint on the
point of maximum absolute radial velocity. We ran the Lucy-Sweeney
test to determine if the derived small eccentricities in our orbital
solutions of the remaining datasets were significant
\citep{Luc71}. The probability of small eccentricity values arising by
chance were $>5\%$ in all cases. Thus, we fixed the eccentricity to
zero (circular orbits) in our subsequent modelling with
\textsc{Exofast}, which allowed us to determine the velocity offset
values for each dataset. These offsets are listed in the notes of
Table~\ref{tab:rvdata}, and were subsequently used to place all the
RVs onto the same zero-point velocity.

Prompted by the report of a $1.64$ m~s$^{-1}$~yr$^{-1}$ trend in the
RVs of 51 Peg by \citet{But06} and scatter in the discovery RVs
reported by \citet{May95}, we assessed the significance of long-term
linear trends in our circular orbit solutions. We found that the
earliest Lick RVs (Lick13), spanning 791 days, supported a
$-1.64_{-1.10}^{+1.17}$ m s$^{-1}$ yr$^{-1}$, in agreement with
\citet{But06}. However, the remaining Lick datasets, spanning 1175
days and 3354 days, respectively, returned non-significant linear
trends of $-0.58_{-0.88}^{+0.84}$ m s$^{-1}$ yr$^{-1}$ and
$0.029_{-0.29}^{+0.28}$ m s$^{-1}$ yr$^{-1}$, respectively. The ELODIE
observations, including concurrent RVs with the Lick13 measurements,
however suggest a much smaller trend ($-0.15_{-0.40}^{+0.37}$ m
s$^{-1}$ yr$^{-1}$), in better agreement with the more recent HIRES
RVs ($-0.33_{-0.19}{+0.19}$ m s$^{-1}$ yr$^{-1}$). It is possible that
the ensemble RV data for 51 Peg probe the turn over point of an RV
curve for a long-period companion. However, it is beyond the scope of
this paper to search for such a companion. To obtain our final orbital
solution for 51 Peg b, we place all the RVs onto the zero-point
described above, and model the system as a circular orbit, with no
long-term linear trend. Our orbital solution is given in
Table~\ref{tab:orbital} and shown in Figure~\ref{fig:rv}. The
correlation between the parameters in the fit are shown in the
Appendix along with their covariance values. We note that the
additional RV scatter from any long-term companion does not cause
prohibitively large error bars in the ephemeris for our later
analysis.

\begin{deluxetable*}{lll}
  \tablecaption{Updated orbital solution and planet properties for 51 Peg b}
  \tablehead{\colhead{Parameter}&\colhead{Units}&\colhead{Value}}
  \startdata \sidehead{Stellar Parameters:}
  ~~~$M_{\star}$\dotfill &Mass ($\msun$)\dotfill & $1.100\pm0.066$\\
  ~~~$R_{\star}$\dotfill &Radius ($\rsun$)\dotfill & $1.020_{-0.079}^{+0.084}$\\
  ~~~$L_{\star}$\dotfill &Luminosity ($\lsun$)\dotfill & $1.05_{-0.24}^{+0.32}$\\
  ~~~$\rho_\star$\dotfill &Density (cgs)\dotfill & $1.46_{-0.27}^{+0.34}$\\
  ~~~$\log(g_\star)$\dotfill &Surface gravity (cgs)\dotfill & $4.452_{-0.059}^{+0.061}$\\
  ~~~$\teff$\dotfill &Effective temperature (K)\dotfill & $5790\pm230$\\
  ~~~$\feh$\dotfill &Metalicity\dotfill & $0.198\pm0.029$\\
  \sidehead{Planetary Parameters:}
  ~~~$P$\dotfill &Period (days)\dotfill & $4.2307869_{-0.0000046}^{+0.0000045}$\\
  ~~~$a$\dotfill &Semi-major axis (AU)\dotfill & $0.0528_{-0.0011}^{+0.0010}$\\
  ~~~$T_{eq}^{a}$\dotfill &Equilibrium Temperature (K)\dotfill & $1226_{-69}^{+72}$\\
  ~~~$\fave$\dotfill &Incident flux (\fluxcgs)\dotfill & $0.51_{-0.11}^{+0.13}$\\
  \sidehead{RV Parameters:}
  ~~~$K_{\star}$\dotfill &RV semi-amplitude (m/s)\dotfill & $54.93\pm0.18$\\
  ~~~$M_{\rm P}\sin i$\dotfill &Minimum mass (\mj)\dotfill & $0.466\pm0.019$\\
  ~~~$M_{\rm P}/M_{\star}$\dotfill &Mass ratio\dotfill & $0.0004043_{-0.0000080}^{+0.0000085}$\\
  ~~~$\gamma_{\rm zp}$\dotfill &Residual zero-point offset$^{b}$ (m/s)\dotfill & $0.032_{-0.077}^{+0.076}$\\
  ~~~$T_C$\dotfill &Time of conjunction (\bjdtdb)\dotfill & $2456326.9314\pm0.0010$\\
  ~~~$P_{T,G}$\dotfill &A priori transit probability\dotfill & $0.092_{-0.013}^{+0.014}$\\
  ~~~$M_{p}$\dotfill &Planet mass (\mj)\dotfill & $0.476^{+0.032}_{-0.031}$\\
  ~~~$i$\dotfill &Inclination ($^{\circ}$)\dotfill & $70-90$\\

  \enddata
  \tablecomments{The stellar RV parameters were derived with \textsc{Exofast},
    based on priors for $\log(g)$, $T_{\rm eff}$, [Fe/H], and
    $\log_{10}(P)$ whose values are noted in
        the main text of Section~\ref{sec:orbit}. The eccentricity was
        fixed to $e=0$ and we did not allow for a long-term linear
        trend. The true planet mass and its range of inclination were
        derived from the CRIRES spectra as detailed
        in Section~\ref{sec:doubleline}.}
  \tablenotetext{a}{$T_{eq}$
        was calculated assuming zero Bond albedo, $A_{B}=0,$ and
        perfect redistribution of incident flux following
        \citep{Hans07}.}
  \tablenotetext{b}{This is the residual
        scatter around the zero-point determined from the independent
        orbital fits. It is consistent with zero within the $1\sigma$
        error bars, and we note that the velocity resolution of the
    CRIRES pixels ($1.5$ km~s$^{-1}$) is insensitive to this small
    discrepancy. We adopt a literature systemic velocity of $-33.25$
    km~s$^{-1}$ \citep{Brog13}.}
  \label{tab:orbital}
\end{deluxetable*}

\begin{figure*}
  \centering
  \includegraphics[width=\textwidth]{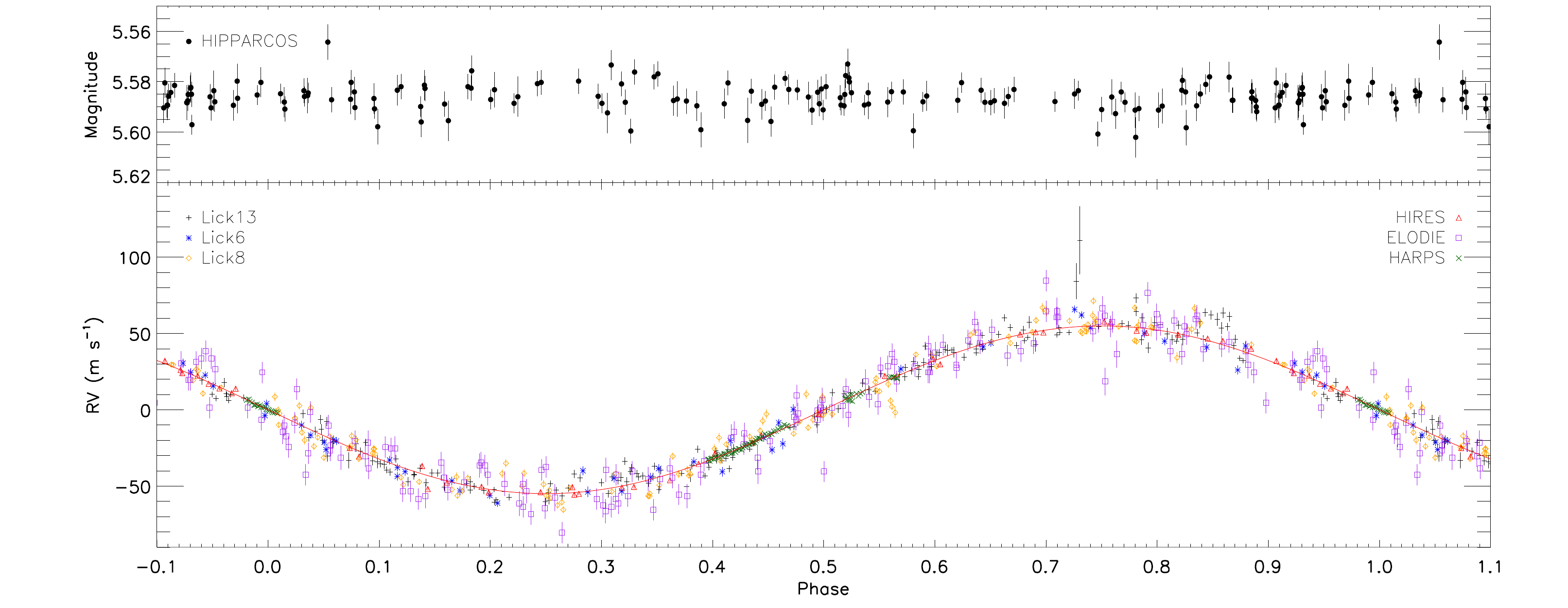}
  \caption{\textbf{Top:} Photometric monitoring of 51 Peg from
    Hipparcos, phase-folded on the orbital period of 51 Peg b. The
    x-axis is plotted such that transit would occur at phase=0,
    respectively, if present in the light curve. \textbf{Bottom:}
    Radial velocity curve for 51 Peg with our updated orbital solution
    (red solid line). The reduced $\chi^{2}$ of the best fitting model
    displayed here was 6.9, thus \textsc{Exofast} rescaled all of the
    RV uncertainties by a multiplicative factor of 2.6 to achieve
    $\chi^{2}_{\nu}=1$ and counteract any underestimate of the RV
    uncertainties. The errors on the derived parameters in
    Table~\ref{tab:orbital} reflect these increased RV
    uncertainties. The error bars in this plot have not been rescaled
    from the original literature values given in
    Table~\ref{tab:rvdata}.}
  \label{fig:rv}
\end{figure*}

For reference, we also show in the top panel of Figure~\ref{fig:rv}
photometric monitoring data obtained with Hipparcos \citep{hip97}
which were extracted from the NASA Exoplanet Archive. The data
were obtained between November 1989 and December 1992. The plotted
data only include measurements with a quality flag of zero, and are
zoomed such that variations at the few percent level could be clearly
seen. There are no transit events within the scatter of the data when
folded on the orbital period of 51 Peg b. This is entirely consistent
with more recent ground- and space-based photometric monitoring of 51
Peg, which also report no transit events corresponding to Earth-size
planets or larger at the orbital period of 51 Peg b
\citep{May95IAU,Gui95,Hen97,Walk06}. This places an upper limit on the
orbital inclination angle, which is discussed in
Section~\ref{sec:doubleline}.

With the orbital solution updated, we re-ran the cross-correlation
analysis using our new ephemeris for 51 Peg b. This resulted in the
strongest cross-correlation signal appearing $-13.3\pm2$ km~s$^{-1}$
from the known systemic velocity of the host star i.e. even further
offset than when using the \citet{But06} orbital calculation. In order
to match the signal to the know systemic velocity, we must invoke a
phase shift of $\Delta\phi=0.1$, or equivalently $\Delta T_{C}=0.07$,
which is significantly larger than our $1\sigma$ error on $T_{C}$. One
therefore might conclude that the strongest cross-correlation signal
is not associated with 51 Peg b. However, given that we see a strong
signal at almost identical offsets across multiple data sets
(i.e. those reported here and at $2.3~\mu$m by \citealt{Brog13}),
which targeted different wavelength regimes and different molecules,
and underwent different data processing techniques, the case of
association with the star remains plausible. While it is possible that
additional long period companions in the 51 Peg system could be
affecting the orbit of 51 Peg b, it seems unlikely that such a
companion could induce the magnitude of shift in phase we have
measured. We instead note that constraints on certain parameters in
the orbital solution using solely the stellar RVs are not strong,
especially for the argument of periastron ($\omega_{\star}$). Using
our orbital solution in Table~\ref{tab:orbital}, but allowing a small
eccentricity ($e=0.001$), which cannot be constrained by the existing
stellar RV data, we find that an offset in $\omega_{\star}$ of just
$6^{\circ}$ from the standard definition ($\omega_{\star}=90^{\circ}$)
aligns the strongest cross-correlation signal with the known systemic
velocity of the host star. This is considerably smaller than the
typical $\omega_{\star}$ uncertainties reported in the literature from
stellar RV measurements. For example, using the Exoplanet Data
Explorer\footnote{http://exoplanets.org} \citep{Han14}, we find that
for similar, non-transiting hot Jupiter systems ($P<10$ days) with
small orbital eccentricity ($e<0.1$), the smallest reported error on
$\omega_{\star}$ is $\pm11^{\circ}$, and the rest are all larger than
$36^{\circ}$. Given that such small changes in the orbital solution
can result in alignment of the star and planet systemic velocities, we
conclude that the most likely scenario arising from our data is that
strongest cross-correlation signal is associated with the 51 Peg
system, and thus of planetary nature. However, for simplicity, we
adopt a phase shift of $\Delta\phi=0.1$ in the circular orbital
solution, as noted above, to align the signal, rather than modify the
eccentricity and $\omega_{\star}$.

We further note that recent studies have highlighted that the
cross-correlation of water models from different molecular line list
databases can result in a velocity offset due to mismatching lines
\citep{Brog17}. In the case of 51 Peg b, the velocity shift is seen
with both water and CO models, where the latter molecule has a very
robust line list. Consequently, we think it unlikely that water line
lists are causing the velocity shift we observe in 51 Peg b, but we
highlight it as a potential issue for other future studies of
exoplanet atmospheres at high spectral resolution.

We further conducted our cross-correlation analysis using other
molecular templates containing signatures of CH$_{4}$, CO$_{2}$ for
similar grid parameters, but we found no significant ($>3\sigma$)
signal from these molecules that would indicate their presence at the
abundances probed by our model grid in the atmosphere of 51 Peg b (see
Section~\ref{sec:atmos_props} for a discussion on the possible causes
for these non-detections).

\subsection{Determining the significance of the cross-correlation
  detection}\label{sec:significance}
\begin{figure}
\centering
\includegraphics[width=0.5\textwidth]{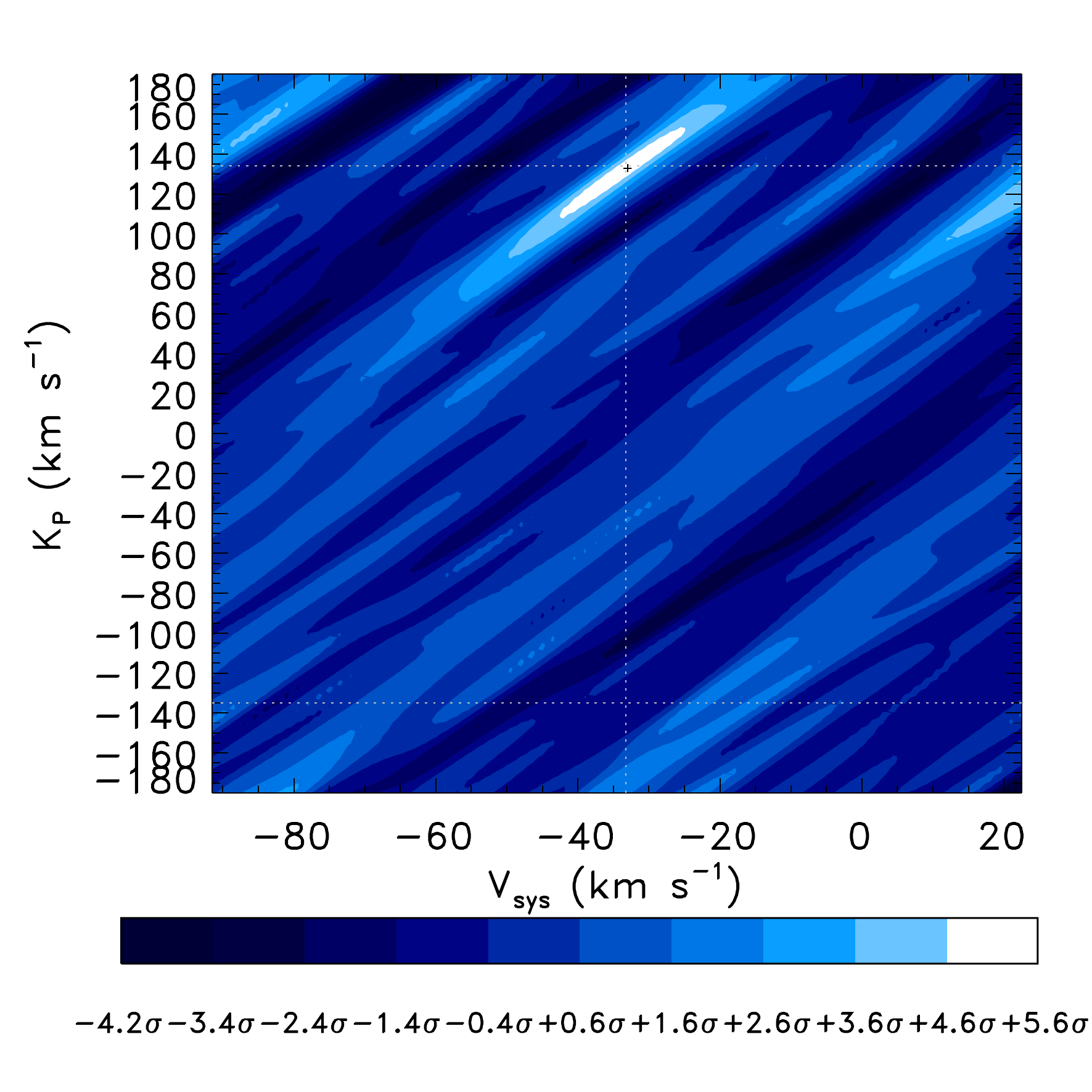}
\caption{Significance values derived from T-test for the best-matching
  template. The black plus sign marks the peak significance,
  $5.6\sigma$, located at $V_{\rm sys}=-33\pm2$ km~s$^{-1}$ and
  $K_{P}=133^{+4.3}_{-3.5}$ km~s$^{-1}$, where the errors are the
  maximal extent of the white one sigma error contour. The dashed
  white lines mark the known systemic velocity ($V_{\rm sys}=-33.25$
  km~s$^{-1}$), and the tentative reported value of
  $K_{P}=134.1\pm1.8$ km~s$^{-1}$ by \citet{Brog13}. The peak
  significance and its $1\sigma$ error contour are coincident with
  these literature values. A dashed white horizontal line at
  $K_{p}=-134.1$ km/s highlights that there is no matching signal in
  negative velocity space, adding as an additional sanity check
  against spurious signals. }
\label{fig:ttest}
\end{figure}

In the final CCF matrix the planet signal appears as a dark diagonal
trail that, in this case, is blue-shifting across the matrix. The
trail is a small section of the planet RV curve and its slope
determines the planet velocity. For visual purposes only, the bottom
left panel of Figure~\ref{fig:trail} shows the CCF matrix after the
best-matching template was injected into the observed spectra, before
the removal of telluric contamination, at $7\times$ its nominal value
at the detected velocity of the planet. When shifted into the planet's
rest frame velocity, the trail becomes vertically aligned, as shown in
the middle panels of Figure~\ref{fig:trail}. The size of the shift to
vertically align the signal in each CCF is related to the RV
semi-amplitude of the planet ($K_{P}$). Although \citet{Brog13}
reported a tentative value of $K_{P}=134.1\pm1.8$ km~s$^{-1}$ for 51
Peg b, we opted to search for the planet signal over a wide range of
$K_{P}$ values to act as a blind search that would independently
confirm this value. We therefore aligned CCFs for a range of $K_{P}$
values, from $20\leq K_{P}(\rm km~s^{-1})\leq180$ in steps of $1$
km~s$^{-1}$. To determine when the trail was vertically aligned,
i.e. in the planet rest frame, thus yielding the value of $K_{P}$, we
performed a Welch T-test on each aligned CCF matrix. This statistic
compared the distribution of a 3-pixel wide sliding column of
`in-trail' CCFs values in the aligned matrix to the distribution of
those outside it (`out-of-trail'), and determined the probability that
they were drawn from the same parent distribution. The sliding of the
column allows for the location of $V_{\rm sys}$, as well as
$K_{P}$. These probabilities were converted into $\sigma$-values,
using the \textit{erf} \textsc{IDL} function, and are displayed as a
matrix in Figure~\ref{fig:ttest}, where $V_{\rm sys}$ denotes the
systemic velocity of the central column of the in-trail data. The most
discrepant in- and out-of-trail distributions deviated by $5.6\sigma$
and corresponded to $V_{\rm sys}=-33\pm2$ km~s$^{-1}$, which is
coincident with the known systemic velocity of the host star, and at
$K_{P}=133^{+4.3}_{-3.5}$ km~s$^{-1}$, as marked by the black cross in
Figure~\ref{fig:ttest}. The error bars are the $1\sigma$ marginalised
uncertainties, although it is clear they are correlated based on the
ellipsoidal shape of the $1\sigma$ error contour shown in white in
Figure~\ref{fig:ttest}. This is expected, because as we approach the
correct $K_{P}$ for the planet when aligning the CCFs, the detection
strength will increase. If the observations had occurred before
superior conjunction, the contour would be slanted in the opposite
direction. Combining multiple nights of similar data, but spanning
different phase ranges, would significantly constrain the $1\sigma$
contour (see e.g. \citealt{Brog12}). Following \citet{Brog14}, in
Figure~\ref{fig:ttest} we also explore negative orbital velocities for
the planet. Negative velocity implies a retrograde orbit which we know
is not true for 51 Peg b.  However, any residual correlated noise that
could produce a false positive would interact with the model spectrum
in a similar way in both positive and negative velocity space, with a
false positive appearing as a mirror image of the positive signal in
the $K_{p}<0$ km/s plane. The lack of significant negative velocity
signals in Figure~\ref{fig:ttest} therefore serve to highlight the
robustness of the positive velocity detection.

A final step in our analysis was to determine which other model water
templates also provided a detection within $1\sigma$ of the peak
detection significance of $5.6\sigma$ and thus also adequately
describe the observations. Non-inverted models with shallow
temperature gradients (i.e. $t_{1}=1000$ K and $t_{2}=500$ K) returned
comparatively lower significance values between $2.4-4.5\sigma$. All
remaining non-inverted models in the grid, which span the full range
of $p_{1}$, $p_{2}$, and VMR$_{\rm H2O}$ that we explored, were within
$1\sigma$ of our peak detection significance. However, we found that
models in our grid with a temperature inversion (i.e. $t_{2}=1500$ K)
were all negatively correlated with the observed planet spectrum. This
means that the emission lines in the model temperature inversion
spectra correlated negatively with the absorption lines in the
observed planetary spectrum. Consequently, we confidently rule out the
inverted models in our grid as good descriptors of the observations.

\subsection{Removal of degraded spectra}\label{sec:degraded}
\begin{figure}
\centering
\includegraphics[width=0.5\textwidth]{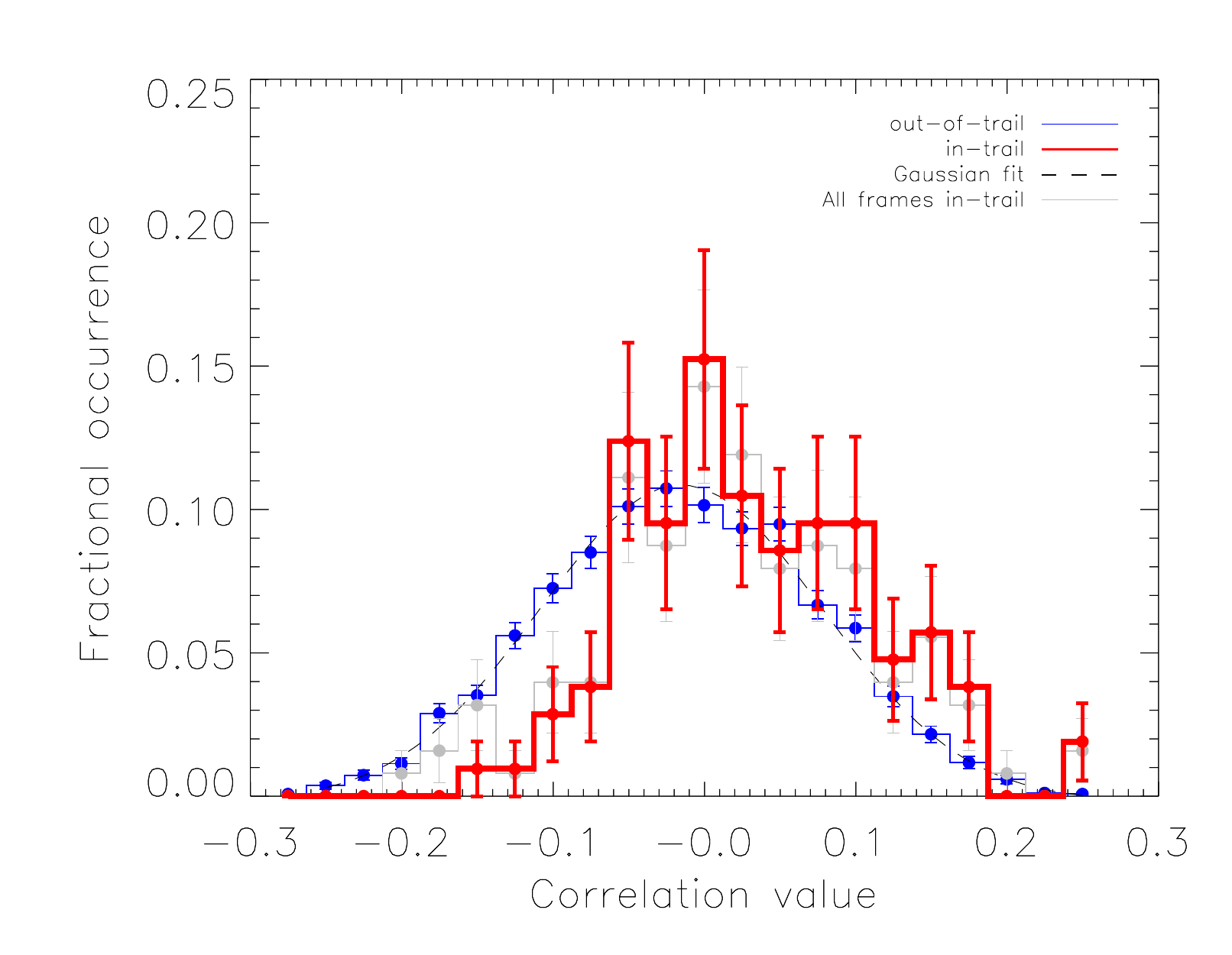}\\
\caption{Histogram showing the difference in aligned CCF
  distributions. The `in-trail' distribution corresponds to a 3-pixel
  wide column in the aligned CCF matrix positioned at
  $V_{\rm sys}=-33$ km~s$^{-1}$, for $K_{P}=133$ km~s$^{-1}$,
  corresponding to the matrix in the upper-right panel of
  Figure~\ref{fig:trail}. The `out-of-trail' distribution contains the
  remaining CCF values outside this column.  The dashed line indicates
  that the out-of-trail CCFs follow a Gaussian distribution. There is
  a noticeable difference in the mean of the in- and out-of-trail
  distributions, which is quantified by the T-test in
  Figure~\ref{fig:ttest}. The grey line shows the in-trail
  distribution when including the degraded data frames (see
  Section~\ref{sec:sysrem}). Note how the grey bins are noticeably
  lower than the red bins for positive correlation values and vice
  versa for the negative bins. The inclusion of these frames in the
  cross-correlation analysis causes a reduction of more than $1\sigma$
  in the difference between the means of the in- and out-of-trail
  distributions. We adopt the red in-trail distribution (i.e. with the
  degraded frames removed) for the remainder of our analysis.}
\label{fig:histogram}
\end{figure}

A histogram of the corresponding in- and out-of-trail distributions
are shown in Figure~\ref{fig:histogram}. Note that there are two
in-trail distributions shown. The grey one includes all of the
observed spectra, while the other (red) one contains only a subset as
described below. The latter was used in our analysis described in
Section~\ref{sec:significance}.

We note that in the model-injected CCFs, the trail fades between
frames 18-24 (see bottom right panel of Figure~\ref{fig:trail}),
suggesting a fundamental degradation of the observations, and
corresponds to the period of unstable seeing noted in
Section~\ref{sec:sysrem}. These CCFs act to scramble the distribution
of the in-trail signal. Their inclusion in the T test results in
$>1\sigma$ decrease in the detection significance ($4.3\sigma$)
compared to when they are excluded (see the grey histogram in
Figure~\ref{fig:histogram} and note how it shifts back towards the
out-of-trail distribution). We have explored multiple reasons for why
these spectra are degraded, as shown in Section~\ref{sec:sysrem}. We
initially attempted to weight the CCFs by the physical parameters
displayed in Figure~\ref{fig:physical}. However, only excessively
large weightings gave an improvement, resulting in only a few CCFs
dominating the final signal. Given that the injected trail also fades,
we conclude that these spectra are not useful. We suspect a
combination of poor atmospheric conditions and thus possible slit
losses are responsible for the degradation. Indeed,
Figure~\ref{fig:physical} shows that frames with the worse seeing
required drift corrections on top of the temperature-induced trend,
and that the flux received by the AO system is considerably lower
during these times.  Consequently, we chose to exclude three frames
either side of the seeing spike, with the central frame being the
first one with the increased exposure time. We note that this may
exclude some planetary signal, thus our detection significance is
conservative. The results of Section~\ref{sec:significance} therefore
include only $35$ of the observed $42$ spectra.

\section{Discussion}\label{sec:discussion}
\subsection{51 Peg Ab: A double-lined spectroscopic binary system}\label{sec:doubleline}
The detection of molecular features in the atmosphere of 51 Peg b
indicates that the 51 Peg Ab system is a double-lined spectroscopic
binary. Our measured $K_{P}$ of 51 Peg b can be combined with
$K_{\star}$ and the system mass ratio determined from precision
stellar RV measurements to determine the mass of 51 Peg b, independent
of its inclination, via
$\frac{K_{\star}}{K_p}=\frac{M_{p}}{M_{\star}}$. Using the stellar
properties given in Table~\ref{tab:orbital}, the measured true mass of
51 Peg b from our observations is
$M_{p}=0.476^{+0.032}_{-0.031}M_{\rm J}$, placing it firmly in the
planetary mass range, and laying to rest any lingering doubts on the
true nature of the first reported exoplanet orbiting a Solar-like
star. The planet's mass places it at the boundary between Jovian and
Neptunian worlds, according to recent work by \citet{Che16}. They
found data-driven evidence for a break in the power-law relationship
between mass and radius for gaseous worlds at
$M_{P}=0.41\pm0.07M_{J}$. This tipping point can be physically
interpreted as the mass at which any further accretion of gas into the
outer layers of a Neptunian atmosphere overcomes the barrier for
self-compression by gravity, leading to Jovians with slightly smaller
radii. Using the mass-radius relations calculated by \citet{Che16}, we
predict that the radius of 51 Peg b should be between
$R_{p}=1.01-1.32R_{\rm J}$, which includes a $14.6\%$ dispersion in
the mass-radius relation. We note that this is considerably smaller
than the $R_{p}=1.9R_{\rm J}$ suggested by recent attempts to detect
the reflected light from 51 Peg b at optical wavelengths
\citep{Mart15}, and we discuss this further in
Section~\ref{sec:reflect}.

Now that we have measured $K_{P}$ for 51 Peg b, we can also solve for
its orbital inclination using Kepler's third law, such that
$PK_{P}/2\pi a=K_{P}/V_{P}=\sin(i)=0.977^{0.038}_{0.032}$ (assuming
$K_{\star}<<K_{P}$, where $V_{P}$ is the orbital velocity of the
planet). This yields an inclination of $i=78^{\circ}$, but the
$1\sigma$ uncertainties permit values within the range
$70<i(^{\circ})<90$. The geometric probability of 51 Peg b transiting
its host star is $P(\rm transit)=R_{\star}/a=9\%$; however, the lack
of transits in photometric monitoring of 51 Peg b (see
Figure~\ref{fig:rv}) place an upper limit on the inclination angle at
$i<82.2^{\circ}$, as described by \citet{Brog13}.  This is also
consistent with estimates of the inclination of the stellar rotation
axis at
$70^{\circ}$$_{-30}^{+11}$ \citep{Sim10}, and suggests spin-orbit
alignment.

\subsubsection{Improving ephemerides for non-transiting planets}
Orbital solutions for non-transiting planets are significantly more
uncertain than for transiting planets, such that the epoch of
periastron of non-transiting hot Jupiters can be uncertain to the
order of half a day, as is the case for 51 Peg b, compared to e.g.
$\pm1.3$ seconds for the transiting hot Jupiter HD 189733 b
\citep{Ago10}.

Although we stress that coincidental alignment of the planet and
stellar systemic velocities is a strong indicator that the detected
molecular absorption signal originates from the planet, in the case
reported here we have multiple datasets, targeting uniquely different
molecules at different wavelengths, that show molecular absorption at
a consistently offset $V_{\rm sys}$, thus providing another means to
assign the signal to the planet 51 Peg b.

In fact, all of the non-transiting planets detected directly with
high-resolution spectroscopy ($\tau$ Boo b, 51 Peg b, and HD 179949;
\citealt{Brog12,Brog13,Brog14}, respectively) have required small
$<1\sigma$ phase offsets to the orbital solution from stellar RVs to
place the signal at exactly the systemic velocity. Conversely, those
that transited (HD 209458 b and HD 189773 b;
\citealt{Sne10,deK13,Bir13}) have not needed phase offsets, arguably
due to their transits enabling far better defined ephemerides.

This leads us to the conclusion that combining precision stellar RVs
with a comprehensive set of high spectral resolution planetary RVs, in
a simultaneous manner, should provide strict constraints on all of the
orbital elements of the star-planet system. For example, increasing
$5^{\circ}$ in the argument of periastron introduces a change in the
51 Peg stellar RVs of only a few m~s$^{-1}$, but changes the planet RVs
by several km~s$^{-1}$. The latter is considerably easier to measure.
Such precise constraints on the orbital elements allow a detailed
investigation into the architecture of exoplanetary systems, as well
as a more precise analysis of their circularisation timescales and the
tidal effects that may affect them. However, there remains an argument
for continued precision RV monitoring of the host stars for
non-transiting planets. Not only is this useful for finding additional
companions, but the initial detection of the Doppler-shifting
planetary signal with high-resolution spectroscopy, and thus the
planet's atmospheric characterization, would be greatly aided by
tightly constrained ephemerides.

\subsection{Atmospheric properties of the planet}\label{sec:atmos_props}
\begin{figure*}
\centering
\includegraphics[width=\textwidth]{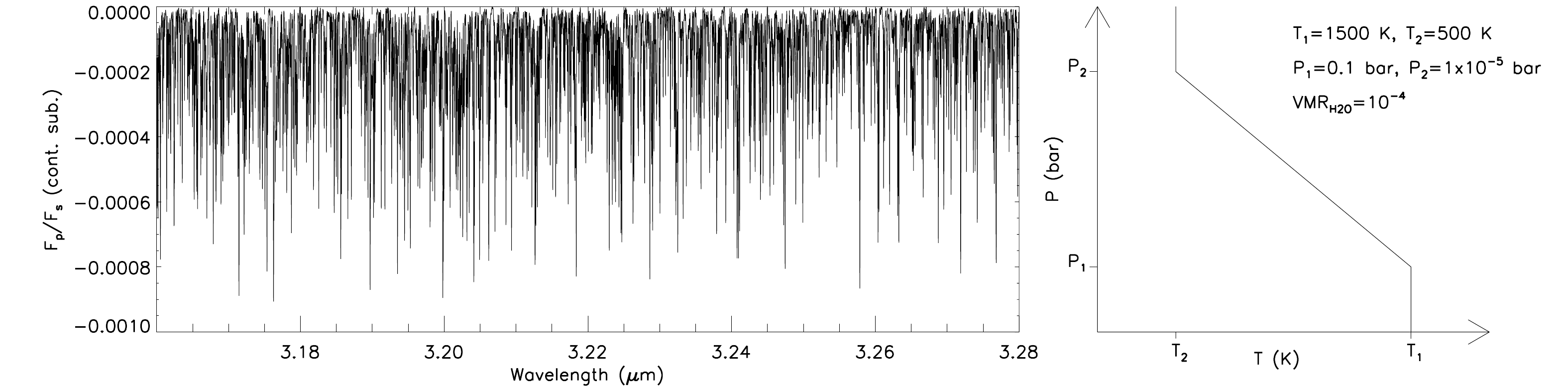}
\caption{{\bf Left:} The best matching H$_{2}$O template, scaled such
  that subtracting it would exactly cancel the signal from the planet
  at $V_{\rm sys}=-33$ km~s$^{-1}$ and $K_{P}=133$ km~s$^{-1}$. The
  template is displayed as a flux ratio, where the planet model was
  divided by a black body spectrum with the same $T_{\rm eff}$ as the
  host star and then had its baseline continuum subtracted and was
  convolved to the CRIRES resolution of $R=100~000$. The deepest lines
  correspond to a relative contrast ratio between the star and planet
  of $1.0\times10^{-3}$ for the unconvolved model, and reduced to
  $0.9\times10^{-3}$ for the convolved model. {\bf Right:} A simple
  schematic displaying the temperature-pressure profile of the
  best-matching template, whose properties are listed in the
  upper-right corner.}
\label{fig:bestmodel}
\end{figure*}

The best-matching model to the planet spectra from our grid contained
molecular features from water only, with a volume mixing ratio of
VMR$_{H_2O}=10^{-4}$, and a $T-P$ profile that steadily reduced in
temperature as altitude increased, from $t_{1}=1500$ K at $p_{1}=0.1$
bar, to $t_{2}=500$K at $p_{2}=1\times10^{-5}$ bar. This model,
convolved to the $R=100~000$ resolution of CRIRES, is shown in
Figure~\ref{fig:bestmodel}, along with a simple schematic of its $T-P$
structure. However, it is important to note that a number of other
models give detection strengths within $1\sigma$ of the best-matching
model at velocities within the $1\sigma$ uncertainties on $K_{P}$ and
$V_{\rm sys}$. This includes models that span the full pressure range
and abundances tested by our grid. The only models in the grid we can
rule out are those that include a temperature inversion within our
probed pressure range, and those with $T_{1}\leq1000$ K. Inversion
layers, in which the best matching templates have emission lines, are
yet to be identified with this method of atmospheric characterization
\citep{Schw15}.

The deepest water lines in the best-matching template have a relative
depth of $1.0\times10^{-3}$. This reduces to $0.9\times10^{-3}$ when
convolved to the resolution of CRIRES. These contrast ratios were
determined by scaling and injecting the negative of the best-matching
template into the observed spectra before the telluric removal
process, at the detected planet velocity parameters, until the peak
detection significance was reduced to zero. The best-matching model
shown in Figure~\ref{fig:bestmodel} required a scaling factor of $3.0$
times its nominal value to completely cancel the planet signal. The
detection of 51 Peg b via ground-based, high-resolution spectroscopy
highlights the method's effectiveness in detecting faint companions
amongst the glare of their much brighter host stars, routinely
reaching contrast ratios between $10^{-4}-10^{-3}$ at angular
separations of $1-3$ milliarcseconds.

No significant detections ($>3\sigma$) were seen when
cross-correlating with models containing carbon dioxide or methane.
While it is likely that the abundances of these molecules are too low
for these data to constrain (see e.g. \citealt{Mad12,Mos13} for
typical abundances of hot Jupiter atmospheres as a function of
temperature and C/O ratio), we do not rule out that incorrect line
positions in the line list used to create the models are responsible
for their non-detection. Hot methane and other line lists are known to
have inaccuracies \citep{Harg15}, particularly at high resolution
\citep{Hoe15}.

\subsubsection{Confirmation of molecular absorption $2.3\mu$m}\label{sec:confirm}
\citet{Brog13} tentatively reported molecular absorption from carbon
monoxide and water in the atmosphere of 51 Peg, using observations
with CRIRES/VLT at $2.3\mu$m in the same month as the observations
presented here. However, \citet{Brog13} found that only two of their
three nights of observations revealed the planet signal, with the
spectral features disappearing when they should have been most
visible, closest to superior conjunction when the largest fraction of
dayside hemisphere is facing towards Earth. They ruled out a secondary
eclipse as the cause of this as the orbital eccentricity would need to
be at least $e = 0.13$, which is strongly excluded by our analysis of
the 51 Peg stellar RVs. \citet{Brog13} also investigated other causes,
including several sources of instrumental instability. However, none
of these could fully account for the loss of the signal. Consequently,
\citet{Brog13} only made a tentative report of the features. Here, we
have detected the orbital motion of 51 Peg b independently, and find
that our derived values for the mass, $K_{P}$, and inclination are
fully consistent with those reported by \citet{Brog13}
($M_{\rm P}=0.46\pm0.02M_{\rm J}$, $K_{P}=134.1\pm1.8$ km~s$^{-1}$,
$V_{\rm sys}=-33.25$ km~s$^{-1}$). We argue that while we used the
same instrument to make our observations, our use of a different
instrument set-up targeting longer wavelengths and a different
molecule, alongside a different technique to remove the contaminating
telluric lines, make it most likely that we have detected the same
astrophysical (i.e. planetary) source, rather than both suffering from
systematics at almost identical $V_{\rm sys}$ and $K_{P}$. We
therefore conclude that we have affirmed the \citet{Brog13} detection
of carbon monoxide in the atmosphere of 51 Peg b.

Following this argument, we compare our best-matching model
atmospheric parameters at $3.2\mu$m with those reported by
\citet{Brog13} at $2.3~\mu$m. The best-matching atmospheric model at
$2.3~\mu$m included VMR$_{CO}=10^{-4}$ and VMR$_{H2O}=3\times10^{-4}$
, for a $T-P$ profile that is $1500$ K at $1$ bar, decreasing without
inversion to 500 K at $10^{-5}$ bar. This is fully consistent with the
$3.2~\mu$m best-matching model within the albeit, large, $1\sigma$
uncertainties. A strong discrepancy would have highlighted that either
clouds were adding to the continuum at the short wavelengths, or that
the continuum probed by each wavelength region was significantly
different. Such discrepancies can be ameliorated by the inclusion of
wide-wavelength high-resolution observations spanning the full
continuum \citep{deK14}, or by the combination of external information
on the true continuum level from low-resolution secondary eclipses or
phase curves.

The observations at $2.3~\mu$m were taken on opposite sides of the
orbit. Therefore, the slanted error contours from each night (similar
to that seen in Figure~\ref{fig:ttest}) cross diagonally thus
narrowing the $1\sigma$ error contour on $K_{P}$ and hence reducing
the $1\sigma$ uncertainties on the planetary mass. Combining the
comparatively small set of $3.2~\mu$m observations with those at
$2.3~\mu$m does not improve the error bars on the orbital properties
from \citet{Brog13}, thus we refrain from including this detailed
analysis in this report. We instead highlight that comparative
datasets taken at optimal wavelengths for simultaneously detecting all
of the other major carbon- and oxygen-bearing molecules (H$_{2}$O,
CH$_{4}$, and CO$_{2}$) could result in more precise constraints on
the relative abundance, expected to be within one order of magnitude
\citep{deK14}.

\subsubsection{Other atmospheric observations of 51 Peg b}\label{sec:reflect}
\begin{figure}
\centering
\includegraphics[width=0.5\textwidth]{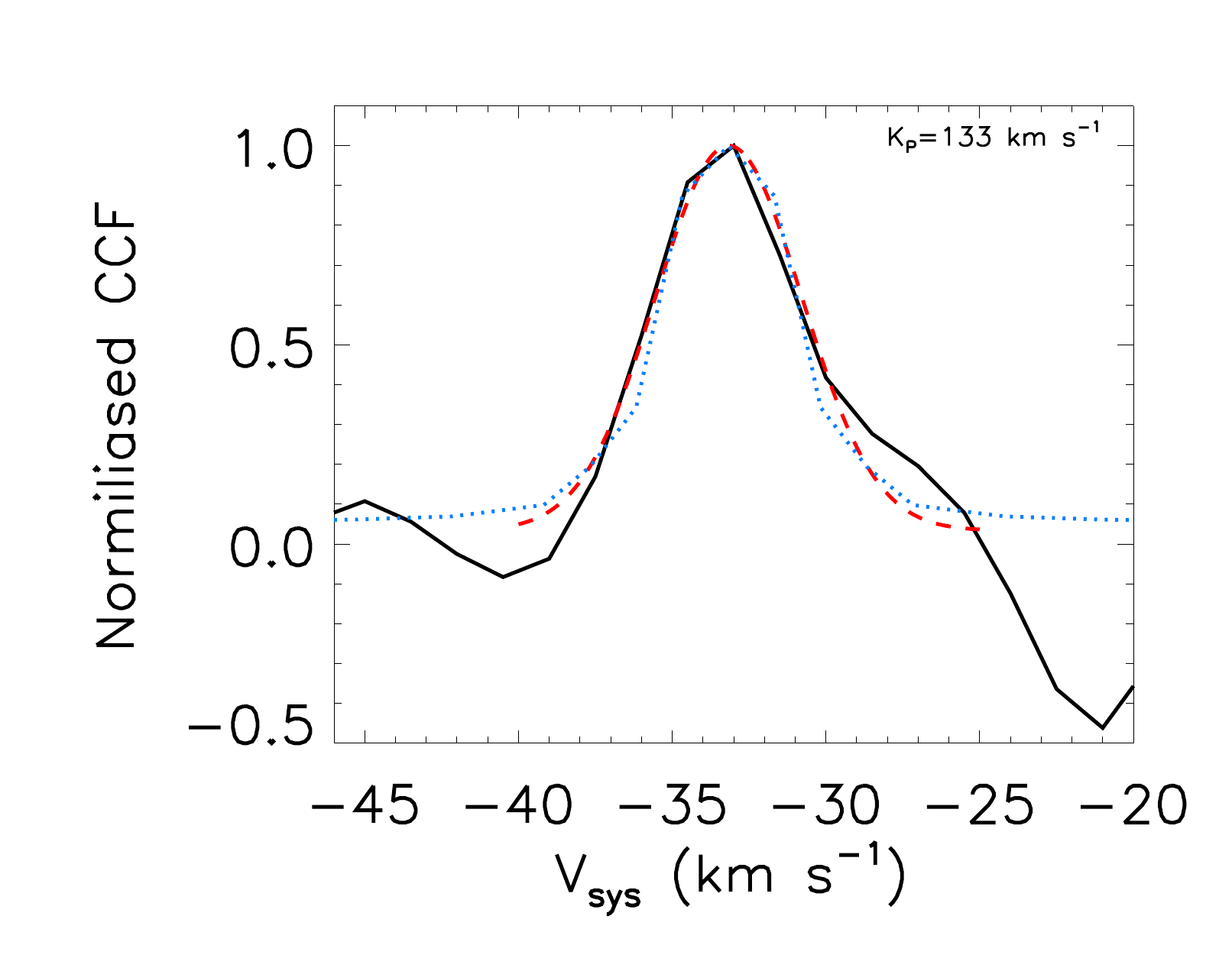}
\caption{Section of the summed and normalised CCF for 51 Peg b (solid
  black line), assuming $K_{P}=133$ km~s$^{-1}$. The dashed red line
  is the best-fitting Gaussian to the CCF centered at
  $V_{\rm sys}=-33$ km~s$^{-1}$. The Gaussian has a FWHM$=5.6$
  km~s$^{-1}$. The dotted blue line is the cross-correlation function
  of the best-matching template with itself after broadening the
  template to match a spectral resolution of $R=100~000$.  }
\label{fig:CCF}
\end{figure}

\citet{Mart15} reported a tentative ($3\sigma$) detection of reflected
light from 51 Peg b using high-resolution ($R=115~000$) spectra
observed with HARPS/ESO-3.6 in 2013, implying a planet-to-star flux
ratio on the order of $F_{p}/F_{s}=10^{-4}$ at optical wavelengths.
They calculated a geometric albedo of $A_{g}=0.5$ under the assumption
of a highly inflated planet at $R_{P}=1.9R_{J}$, while smaller
planetary radii (e.g. $R_{P}=1.2R_{J}$) would require very high
albedos ($A_{g}>1$), inconsistent with a Lambertian sphere (i.e.
isotropic reflection in all directions) implying a strongly
backscattering atmosphere. The CCF derived by \citet{Mart15} for 51
Peg b was also significantly broadened to a FWHM$=22.6\pm3.6$
km~s$^{-1}$, which at face value implies rapid rotation of the
planet's atmosphere at the altitudes (pressures) probed by the optical
spectra. Assuming that 51 Peg b is tidally-locked to its host star,
the expected rotation speed of the planet is
$V_{rot}=2\pi R_{P}/P=1.2-2.3$ km~s$^{-1}$ for $R_{P}=1-1.9R_{J}$.
This is close to or below the resolution of the instrument profile for
both HARPS and CRIRES. We show the CCF from the CRIRES observations at
the detected planet velocity ($K_{P}=133$ km~s$^{-1}$) in
Figure~\ref{fig:CCF}. The CRIRES CCF matches the cross-correlation of
the best-matching template cross-correlated with a version of itself,
after being broadened to match the $R=100~000$ resolution of
CRIRES. Using the measured FWHM$=5.6$ km~s$^{-1}$ of the CRIRES CCF,
combined with the lower limit on the orbital inclination of the system
and assuming the spin axes are aligned, we derive an upper limit to
the rotational broadening of 51 Peg b of $5.8$ km~s$^{-1}$
($P_{rot}>0.9$ days). This result applies to the atmospheric rotation
at the altitude and pressure probed by our infrared observations.
Rather than invoke dynamical arguments e.g. wind sheer, or an exo-ring
system \citep{San15} for the discrepancy with the HARPS optical
measurement of the FWHM, we instead highlight that \citet{Mart15}
urged caution regarding their FWHM measurement, stressing that
injected planet spectra and the host star spectrum both resulted in
over-broadened CCFs in their data analysis, possibly due to
non-Gaussian noise in the data, indicating that the parameters of
their CCFs were strongly affected by the noise for signals at the
$3\sigma$-level.

We also note that the orbital solution used by \citet{Mart15} assumed
a fixed, zero eccentricity, but did not explore the systemic velocity
parameter space, which may have lead to a stronger signal being
detected at the offset $V_{\rm sys}$ we found in
Section~\ref{sec:orbit}. If the optical data can be analysed such that
it does not induce the excess broadening seen in \citet{Mart15}, we
can compare the detected planet $V_{\rm sys}$ at optical and infrared
wavelengths. If the reflected light signal is not offset in
$V_{\rm sys}$, while the infrared remains discrepant, then assuming it
is astrophysical it potentially indicates an offset hot spot in the
non-transiting planet's highly irradiated atmosphere, akin to that
seen in the transiting hot Jupiters HD 189733 b and WASP-43 b
(\citealt{Knu07,Stev14}, respectively), enabling a means of mapping
the atmospheres of close-in non-transiting planets. This would require
small errors on the measured $V_{\rm sys}$, which can be significantly
reduced by observing the planet over a range of orbital phase,
especially before and after superior conjunction. To-date, the best
constrained $V_{\rm sys}$ of a hot Jupiter is for $\tau$ Boo b
\citep{Brog12} with an uncertainty of $\pm0.1$ km/s (an order of
magnitude better than reported here for 51 Peg b). $\tau$ Boo is of
similar brightness to 51 Peg b and required 3 half nights of CRIRES
observations to achieve this precision on $V_{\rm sys}$, a factor of 3
times longer than our 51 Peg b observations. Similar observing
durations would be required in the optical. This suggests that mapping
of non-transiting hot Jupiters could be achieved with current
instrumentation on 4-8 m class telescopes.

New and future instrument, such as ESPRESSO/VLT, and G-CLEF/GMT, will
enable more sensitive probes of the reflected light and the albedo of
even smaller (non-)transiting planets, while newly commissioned,
improved, and upcoming high-resolution infrared spectrographs (e.g.
IGRINS, GIANO, iSHELL, CARMENES, NIRSPEC, and CRIRES+, to name a few,
see \citealt{Cross14} for an extensive list) will independently
confirm the CRIRES infrared detections to date, and probe smaller,
cooler planets.

Finally, we note that dedicated, stable, uninterrupted, and consistent
observing sequences under stable atmospheric conditions best serve the
analysis of exoplanet atmospheres at high spectral resolution.

\section{Conclusions}\label{sec:conclusions}
We have presented a $5.6\sigma$ detection of water molecules in the
atmosphere of the original hot Jupiter, 51 Peg b, providing the first
confirmation that the 51 Peg Ab system is a double-lined spectroscopic
binary. The companion orbits with a radial velocity of
$K_{P}=133^{+4.3}_{-3.5}$ km~s$^{-1}$, and has a measured mass of
$M_{p}=0.476^{+0.032}_{-0.031}M_{\rm J}$, placing it firmly in the
planetary mass regime. Photometric monitoring indicates that the
planet does not transit its host star, but we determine a lower limit
on the inclination to be $i>70^{\circ}$. The reported observations are
sensitive to small ($1\sigma$) changes in the parameters of the orbit
solution derived from stellar RVs. This indicates that combining a
comprehensive set of planet and stellar RVs would significantly
improve the orbital elements. The temperature in the planetary
atmosphere decreases with increasing altitude (non-inverted) over the
pressure ranges probed by these $3.2~\mu$m observations. The detection
of 51 Peg b water spectral features at $3.2~\mu$m adds weight to the
tentative report of CO and H$_{2}$O molecules detected at $2.3~\mu$m
by \citet{Brog13} in two out of three datasets, although we still
cannot explain the lack of detection in their third dataset by
instrumental effects alone. We detected no methane or carbon dioxide
at a significant level ($>3\sigma$) in our observations, indicating a
low abundance, or possibly inaccuracies in the line lists we used to
create the model templates.  Our observations provide an upper limit
to the rotational velocity of this non-transiting planet of
$V_{rot}<5.8$ km~s$^{-1}$, but higher instrument resolution is
required to test if the planet's rotation is tidally-locked to its
host star. Finally, we concluded that further optical observations of
51 Peg b would enable an independent orbital solution in reflected
light, and if this resulted in a significantly different planet
systemic velocity from the infrared observations, that an offset hot
spot in the atmosphere may be needed to explain the infrared
measurements. After 21 years, the detailed nature of 51 Peg b is
beginning to reveal itself, yet it remains an intriguing and extreme
solar system.

\begin{figure}
\centering
  \includegraphics[width=\textwidth, trim=3cm 1cm 2cm 4cm]{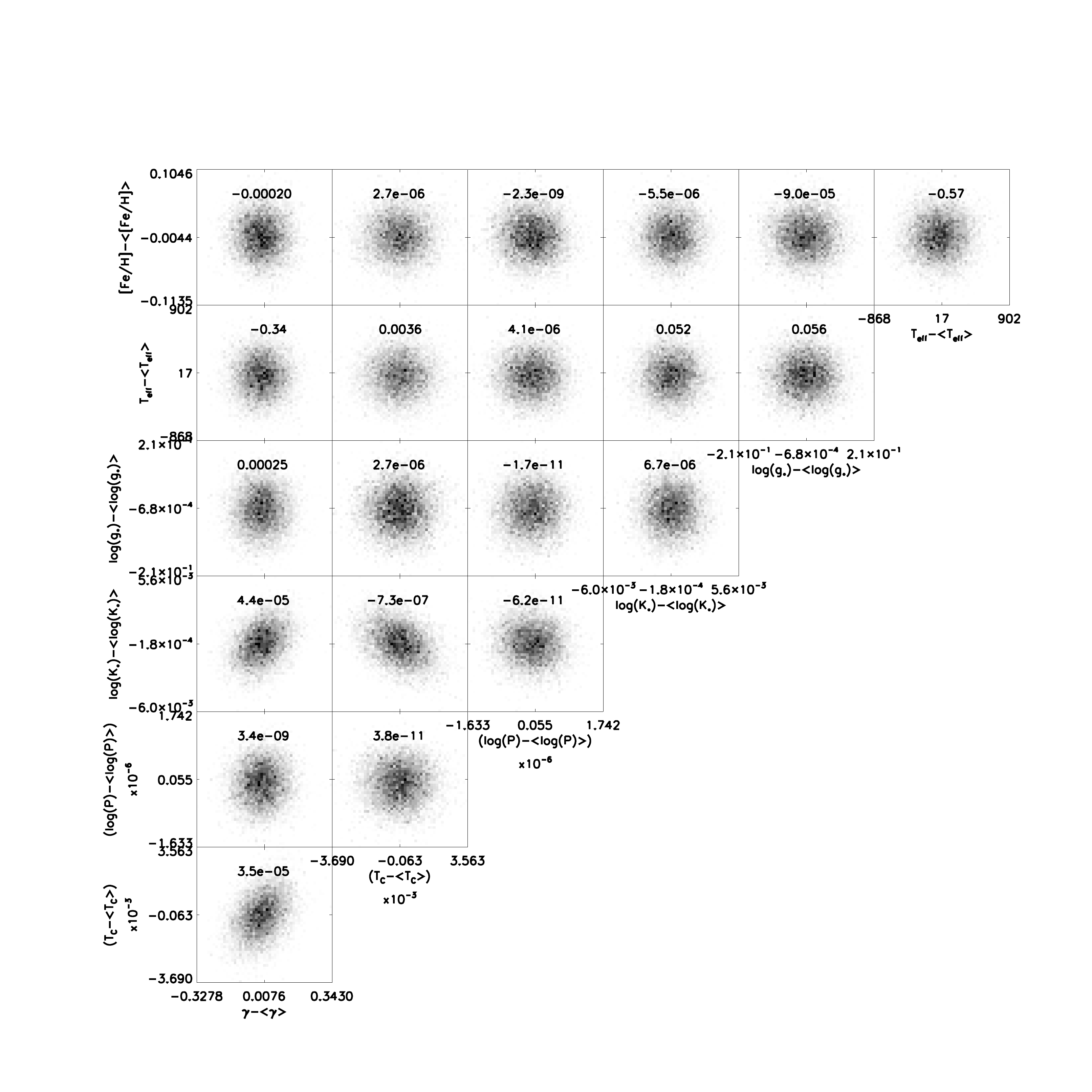}
  \caption{The correlation between parameters in the stellar radial
    velocity fit. The numbers overlaid on each plot give the value of
    the covariance of the parameters. Each parameter is shown with the
    mean of its posterior distribution subtracted. These mean values
    are given at the end of the caption. The parameters shown are as
    follows: residual zero-point offset $\gamma$; time of inferior
    conjunction $T_{C}$; log of the period $\log{P}$; log of the
    stellar radial velocity semi-amplitude $\log(K_{\star})$; log of
    the stellar surface gravity $\log(g)$; effective temperature of
    the star $T_{\rm eff}$; stellar metallicity [Fe/H]; time of
    superior conjunction $T_{S}$; stellar mass $M_{\star}$; stellar
    radius $R_{\star}$; stellar luminosity $L_{\star}$; stellar
    density $\rho_{\star}$; orbital period $P$; semi-major axis $a$;
    the planet equilibrium temperature $T_{\rm eq}$; the incident
    irradiation received by the planet $F_{inc}$; the stellar radial
    velocity semi-amplitude $K_{\star}$; the minimum mass of the
    planet $M_{P}\sin(i)$; the mass ratio $M_{P}/M_{\star}$; total
    duration of any potential transit of the planet $T_{14}$; and the
    a priori transit probability $P_{T,G}$. The mean values subtracted
    from the axis for each parameter in the plots are as follows:
    $\gamma_{\rm mean}$=$0.032$ m/s, $T_{C, \rm mean}$=2456326.9313661
    days, $\log{P}_{\rm mean}$=0.6264212,
    $\log(K_{\star, \rm mean})$=1.740, $\log(g)_{\rm mean}$=4.5 cgs,
    $T_{\rm eff, \rm mean}$=5786 K, [Fe/H]$_{\rm mean}$=0.20,
    $T_{S, \rm mean}$=2456329.047 BJD, $M_{\star, \rm mean}$=1.10
    $M_{\odot}$, $R_{\star, \rm mean}$=1.02 $R_{\odot}$,
    $L_{\star, \rm mean}$=1.08 $L_{\odot}$,
    $\rho_{\star, \rm mean}$=1.49 cgs, $P_{\rm mean}$=4.230787 days,
    $a_{\rm mean}$=0.053 AU, $T_{\rm eq, \rm mean}$=1227 K,
    $F_{inc, \rm mean}$=0.52 $\times10^{9}$ erg s$^{-1}$,
    $K_{\star, \rm mean}$=54.9 m/s, $M_{P}\sin(i)_{\rm mean}$=0.47
    $M_{J}$, $M_{P}/M_{\star, \rm mean}$=0.00040,
    $T_{14, \rm mean}$=0.12 days, $P_{T,G, \rm mean}$=0.09.}
\label{fig:covariances}
\end{figure}

\acknowledgements We would like to extend our gratitude to the
dedicated staff at ESO VLT for all their assistance in acquiring these
data. We thank Jason Eastman for making \textsc{Exofast} freely
available and easy to use, and for some helpful conversations about
RVs. We also thank David Charbonneau, Mercedes Lopez-Morales, Xavier
Dumusque, Matthew Kenworthy, and Stephen Roberts for helpful
discussions on spectroscopy and statistics during the analysis of
these data. We thank our anonymous referee whose insightful comments
helped improve this manuscript. This work is based on observations
collected at the European Organisation for Astronomical Research in
the Southern Hemisphere under ESO programme 186.C-0289(A). This work
was performed in part under contract with the Jet Propulsion
Laboratory (JPL) funded by NASA through the Sagan Fellowship Program
executed by the NASA Exoplanet Science Institute.  Support for this
work was provided in part by NASA through Hubble Fellowship grant
HST-HF2-51336 awarded by the Space Telescope Science Institute, which
is operated by the Association of Universities for Research in
Astronomy, Inc., for NASA, under contract NAS5-26555. This work was
part of the research program VICI 639.043.107, which is financed by
The Netherlands Organisation for Scientific Research (NWO). This
research has made use of the NASA Exoplanet Archive, which is operated
by the California Institute of Technology, under contract with the
National Aeronautics and Space Administration under the Exoplanet
Exploration Program. This research has made use of the Exoplanet Orbit
Database and the Exoplanet Data Explorer at exoplanets.org.
\bibliographystyle{aasjournal}
\bibliography{referencesjlb}{}
\appendix
In Figure~\ref{fig:covariances}, we show the covariance between the
parameters of the stellar radial velocity fit. The covariance value
for each set is given on each correlation plot. Strong correlation is
seen between the typically expected correlated parameters, e.g.
incident flux and the equilibrium temperature of the planet, which
both depend on the stellar radius, density, and surface gravity, or
the times of superior and inferior conjunction. There are no strong
correlations with the orbital period, nor with the residual zero-point
offset. For this work, the RV semi-amplitude $K_{\star}$ and mass
ratio are key parameters for determining the planet's true
mass. $K_{\star}$ shows no strong correlations with other parameters,
while the mass ratio shows strong correlations with the expected
stellar properties, albeit with small uncertainty ranges.

\begin{figure}
\centering
  \includegraphics[width=\textwidth, trim=3cm 1cm 2cm 4cm]{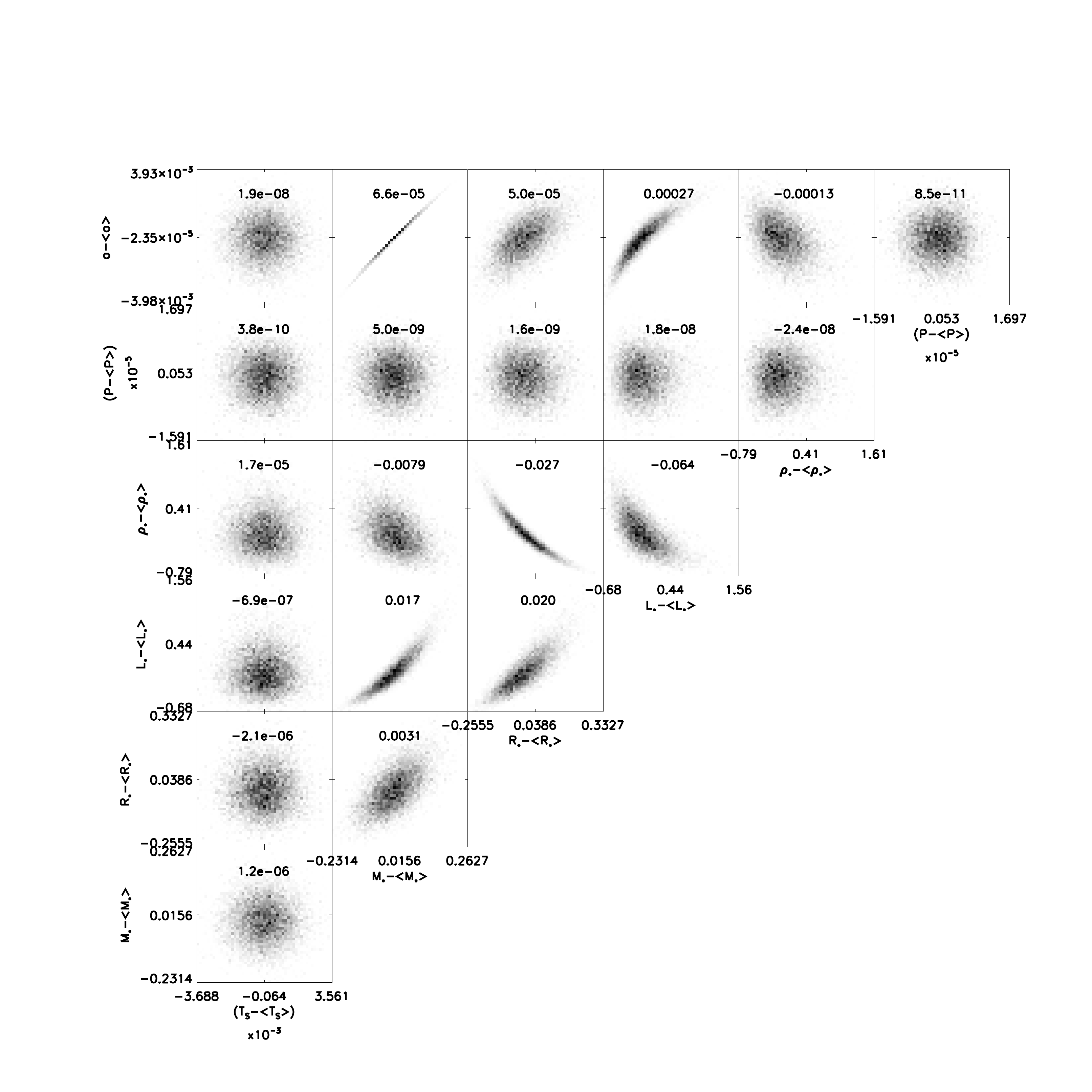}
  \caption{Same as Figure~\ref{fig:covariances} but for additional
    parameters in the radial velocity fit.}
\label{fig:morecovariances}
\end{figure}

\begin{figure}
\centering
  \includegraphics[width=\textwidth, trim=3cm 1cm 2cm 4cm]{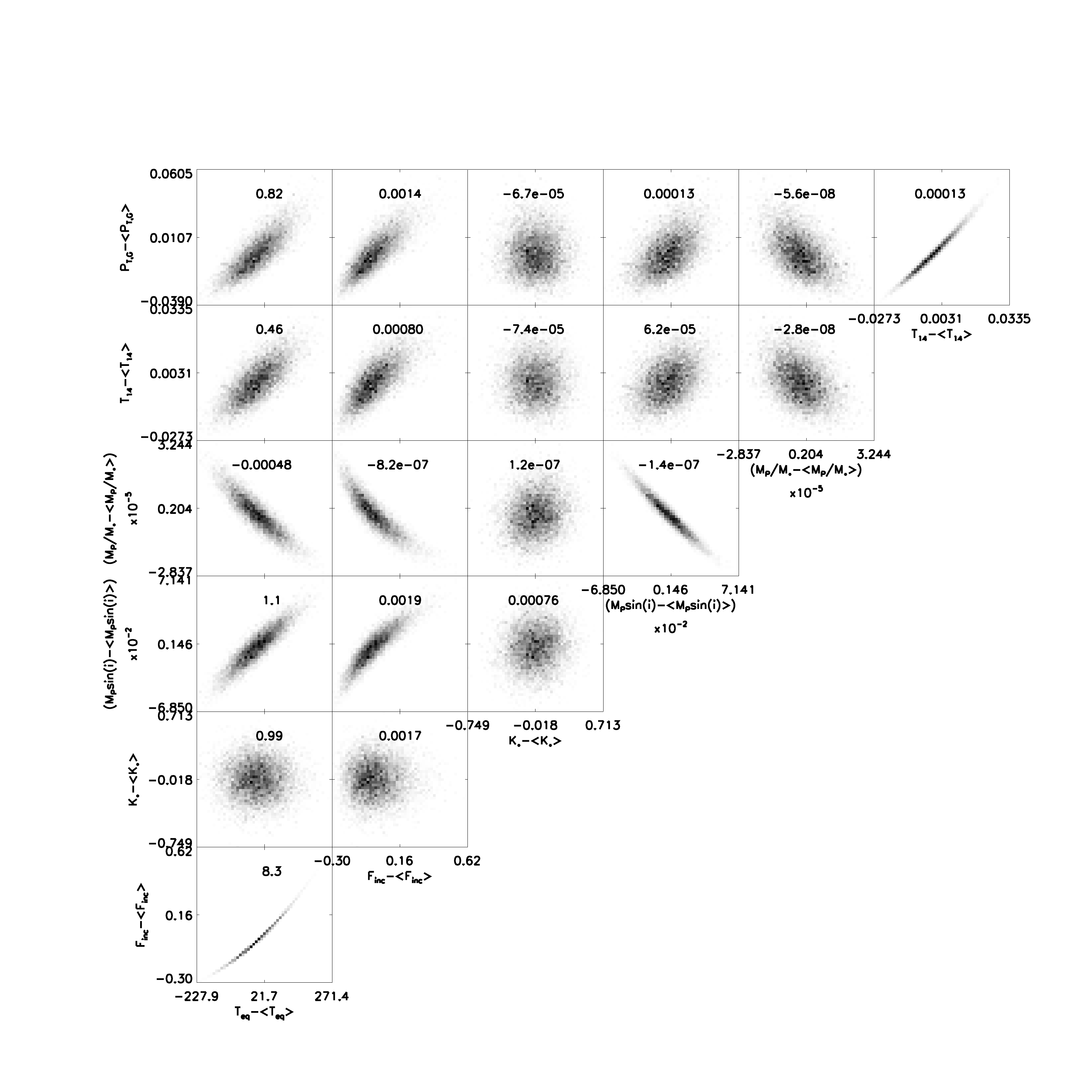}
  \caption{Same as Figure~\ref{fig:covariances}
    and~\ref{fig:morecovariances} but for additional parameters in the
    radial velocity fit.}
\label{fig:lastcovariances}
\end{figure}

\end{document}